\documentclass[conference]{IEEEtran}
\IEEEoverridecommandlockouts
\usepackage{comment}
\usepackage{indentfirst}
\usepackage{graphics}
\usepackage{multirow}
\usepackage{algorithmic}
\usepackage[linesnumbered,ruled,vlined]{algorithm2e}
\usepackage{makecell}
\usepackage{url}
\usepackage{soul}
\usepackage{threeparttable}
\usepackage{footnote}
\usepackage{subfigure}
\usepackage{pgfplots} 
\pgfplotsset{width=8cm, compat=1.6}
\usepgfplotslibrary{groupplots}
\usetikzlibrary{matrix}
\usepgfplotslibrary{external}
\usetikzlibrary{datavisualization}
\usepackage{booktabs}
\usepackage{tikz}
\usepackage{amsthm}
\newtheorem{myDef}{Definition}
\usepackage{array}
\usepackage[ruled,vlined,linesnumbered]{algorithm2e}

\usepackage{color}

\definecolor{eggshell}{rgb}{0.94, 0.92, 0.84}
\definecolor{orange(colorwheel)}{rgb}{1.0, 0.5, 0.0}

\definecolor{shockingpink}{rgb}{0.99, 0.06, 0.75}
\definecolor{beaublue}{rgb}{0.74, 0.83, 0.9}
\definecolor{ballblue}{rgb}{0.13, 0.67, 0.8}

\definecolor{seagreen}{rgb}{0.18, 0.55, 0.34}
\definecolor{springgreen}{rgb}{0.0, 1.0, 0.5}

\definecolor{ufogreen}{rgb}{0.24, 0.82, 0.44}

\definecolor{thistle}{rgb}{0.85, 0.75, 0.85}
\definecolor{lavender(floral)}{rgb}{0.71, 0.49, 0.86}
\definecolor{violet(ryb)}{rgb}{0.53, 0.0, 0.69}

\definecolor{chestnut}{rgb}{0.8, 0.36, 0.36}
\definecolor{fireenginered}{rgb}{0.81, 0.09, 0.13}
\definecolor{goldenyellow}{rgb}{1.0, 0.87, 0.0}
\definecolor{lightkhaki}{rgb}{0.94, 0.9, 0.55}
\definecolor{mediumcandyapplered}{rgb}{0.89, 0.02, 0.17}
\definecolor{oldgold}{rgb}{0.81, 0.71, 0.23}
\definecolor{oldlavender}{rgb}{0.47, 0.41, 0.47}	\definecolor{pastelbrown}{rgb}{0.51, 0.41, 0.33}
\definecolor{persianred}{rgb}{0.8, 0.2, 0.2}
\definecolor{uclagold}{rgb}{1.0, 0.7, 0.0}
\usepackage{cite}
\usepackage{amsmath,amssymb,amsfonts}
\usepackage{graphicx}
\usepackage{textcomp}
\usepackage{xcolor}
\def\BibTeX{{\rm B\kern-.05em{\sc i\kern-.025em b}\kern-.08em
    T\kern-.1667em\lower.7ex\hbox{E}\kern-.125emX}}
\DeclareRobustCommand*{\IEEEauthorrefmark}[1]{%
    \raisebox{0pt}[0pt][0pt]{\textsuperscript{\footnotesize\ensuremath{#1}}}}
    
\begin{document}

\newcommand{\zhu}[1]{\textcolor{blue}{\textbf{/* #1 (Zhu) */}}}

\newcommand{\hui}[1]{\textcolor{red}{\textbf{/* #1 (Hui) */}}}


\title{A Generic Reinforced Explainable Framework\\ with Knowledge Graph for Session-based Recommendation}

\author{\IEEEauthorblockN{Huizi Wu\IEEEauthorrefmark{1}, Hui Fang\IEEEauthorrefmark{2}, Zhu Sun\IEEEauthorrefmark{3}, Cong Geng\IEEEauthorrefmark{1}, Xinyu Kong\IEEEauthorrefmark{4} and Yew-Soon Ong\IEEEauthorrefmark{5}}
\IEEEauthorblockA{\IEEEauthorrefmark{1}\IEEEauthorrefmark{2}Research Institute for Interdisciplinary Sciences, 
School of Information Management and Engineering\\
Shanghai University of Finance and Economics, China\\}
\IEEEauthorblockA{\IEEEauthorrefmark{3}A*STAR,
Institute of High Performance Computing and Centre for Frontier AI Research,
Singapore\\}
\IEEEauthorblockA{\IEEEauthorrefmark{4}Ant Group, China\\}
\IEEEauthorblockA{\IEEEauthorrefmark{5}A*STAR, Centre for Frontier AI Research, Nanyang Technological University, Singapore\\}
\IEEEauthorblockA{\{wuhuizisufe@gmail.com, fang.hui@mail.shufe.edu.cn, sunzhuntu@gmail.com, \\gcong.leslie@gmail.com, xinyu.kxy@antgroup.com, asysong@ntu.edu.sg\}}
\thanks{\IEEEauthorrefmark{2}Hui Fang is corresponding author.}
}

\maketitle

\begin{abstract}
Session-based recommendation (SR) has gained increasing attention in recent years. Quite a great amount of studies have been devoted to designing complex algorithms to improve recommendation performance, where deep learning methods account for the majority.
However, most of these methods are black-box ones and ignore to provide moderate explanations to facilitate users' understanding, which thus might lead to lowered user satisfaction and reduced system revenues. Therefore, in our study, we propose a generic \underline{R}einforced \underline{E}xplainable framework with \underline{K}nowledge graph for \underline{S}ession-based recommendation (i.e., REKS), which strives to improve the existing black-box SR models (denoted as non-explainable ones) with Markov decision process. In particular, we construct a knowledge graph with session behaviors and treat SR models as part of the policy network of Markov decision process. Based on our particularly designed state vector, reward strategy, and loss function, the reinforcement learning (RL)-based framework
not only achieves improved recommendation accuracy, but also provides appropriate explanations at the same time.
Finally, 
we instantiate the REKS in five representative, state-of-the-art SR models (i.e., GRU4REC, NARM, SR-GNN, GCSAN, BERT4REC), whereby extensive experiments towards these methods on four datasets demonstrate the effectiveness of our framework on both recommendation and explanation tasks.
\end{abstract}


\section{Introduction}
\label{sec:intro}
With the rapid development of Internet techniques, session-based recommendation (SR), which focuses on predicting a user's next interested item(s) based on an anonymous session, plays a more and more important role \cite{wang2021survey}. Most existing studies have been devoted to address this task by exploiting complex models like Markov Chain (MC) \cite{le2016modeling} and deep learning (DL) methods, e.g., recurrent neural network (RNN) \cite{li2017neural}, attention mechanism \cite{sun2019bert4rec} and graph neural network (GNN) \cite{wang2020global,pang2022heterogeneous}. Although SR has been widely studied, most of the previous studies focus on improving recommendation accuracy of the system. 
Quite a series of SR methods (denoted as \emph{non-explainable} SR), especially deep learning-based ones, have an unclear internal mechanism and are generally criticized as ``black box''. The lack of transparency and explainability makes users feel difficult to understand the recommendation results and thus be reluctant to accept them. At last, it further leads to lowered user satisfaction towards online experience, and reduced system revenues. Therefore, \emph{explainable} session-based recommender systems, targeting for providing appropriate explanations regarding recommendations besides 
pursuing their accuracy, are in urgent need.

\begin{figure}[htbp]
    \centering\vspace{-2mm}
    \includegraphics[width=7cm]{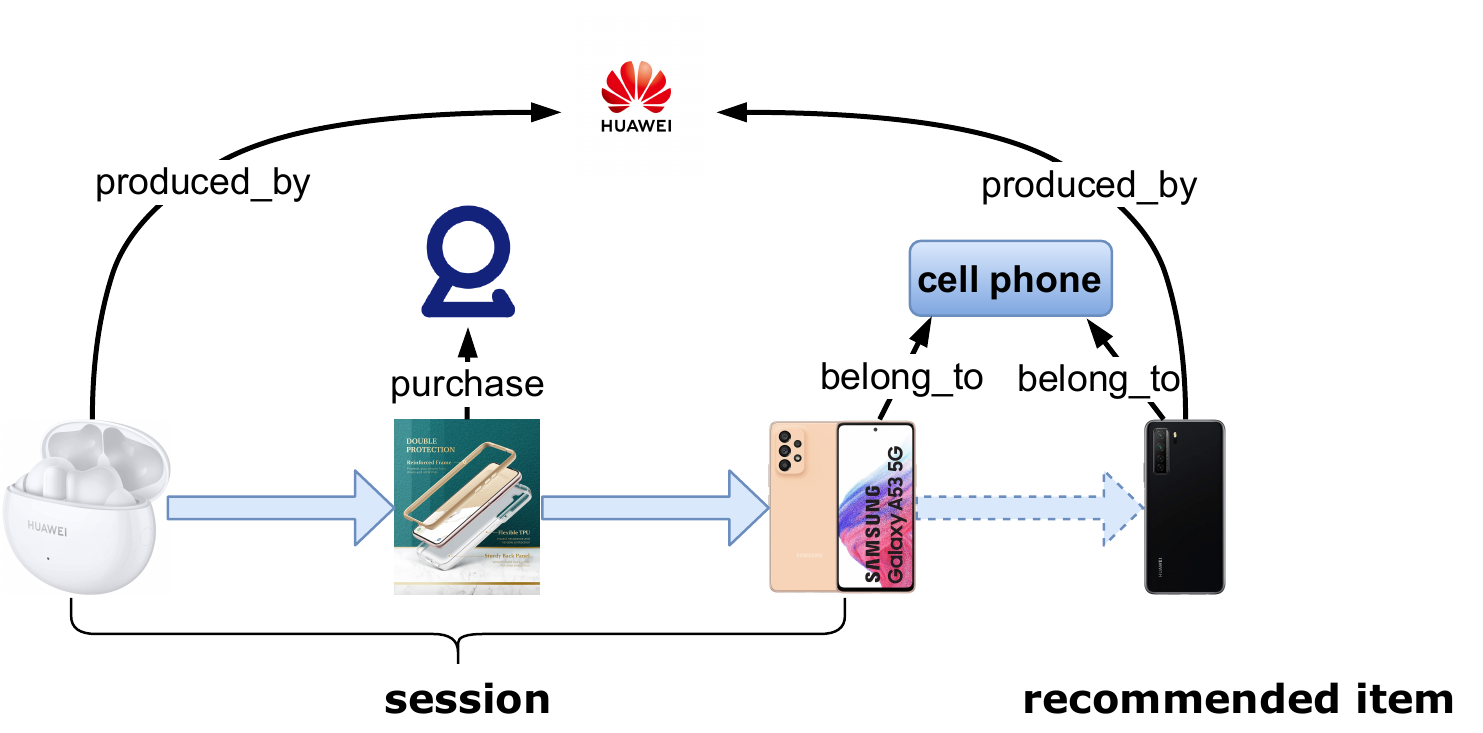}\vspace{-3mm}
    \caption{An Amazon example of a knowledge graph with session information.}\vspace{-2mm}
    \label{fig:kg}
\end{figure}

On the other hand, recently, knowledge graph (KG) is one of the increasingly preferred choices for developing explainable recommendation systems due to its rich information for depicting users/items, as well as their relationships with other entities and attributes~\cite{wang2018ripplenet}. A typical KG consists of entity-relation-entity triplets where under SR scenarios, an entity can be a user or a product, and a relation can be ``purchase'' or ``also\_buy'' between two corresponding entities. Figure \ref{fig:kg} illustrates a KG example from Amazon dataset\footnote{\url{jmcauley.ucsd.edu/data/amazon/links.html}.}. As we can see, there is a session of three products in chronological order. The next recommended item by a certain SR is a phone `produced\_{by}' Huawei probably because the last product of this session is also a phone, and a wireless earphone of Huawei has been historically bought in the session.

KG is firstly adopted in traditional recommendation systems to facilitate the explanation task, which can be concluded into two categories: \emph{post-hoc explainable methods}, and \emph{model-based methods.}
The first category \cite{ai2018learning,wang2018ripplenet} is two-stage, where 
KG embedding-based methods are firstly adopted to learn entity/relation representations from built KGs, which are integrated into recommendation models for providing more accurate results. Then, explanation paths obtained from KGs are generated to give reasonable explanations with regard to corresponding recommendations. 
On the contrary, the second category \cite{wang2019explainable} is end-to-end, since the explanations (mostly also KG-based paths) are obtained simultaneously with recommendation lists, that is, explanation and recommendation tasks are fulfilled together, by using neural symbolic reasoning \cite{xian2020cafe} or reinforcement learning \cite{zhao2020leveraging}. 
The above discussed two types, although generating explanations in different ways, both obtain path-based explanations from KG.

Besides, KG is adopted by sequential recommendation task as well, which also considers user-item interactions in chronological order, but takes user information, and treats a user's all historical interactions as a sequence. This differs from session-based recommendation which mostly considers anonymous sessions\footnote{It should be noted that, in our study, although user information is used to build the knowledge graph, it is merely adopted to better facilitate the explanation task. Besides, without the user information, the proposed generic framework can still work.}.
For example, 
IUM \cite{huang2019explainable} designs an explainable interaction-driven user model to extract semantic paths among user-item pairs and then learns corresponding importance for each path.
The aforementioned methods might be impractical for real-world recommendation systems since it is rather difficult to enumerate all explanatory paths (especially towards longer ones) to measure the importance scores, and then to identify qualified paths for explanation task. 
Besides, they cannot be easily adapted to session-based scenario.


To conclude, there is still little research on explainable SR (except several post-hoc methods \cite{lin2018multi,geng2022path}), whereas quite a great amount of SR algorithms, including DL-based complex ones, have been designed to improve recommendation accuracy. In this view, to go a step further towards explainable SR, it is worthwhile to come up with a \emph{generic path-based explainable framework with KG} to equip existing SR algorithms for the explanation task, instead of designing a brand-new path-based algorithm for explainable SR. 
However, it remains significantly challenging to construct a unified framework on KG for SR, as stated below.

(1) In contrast to the traditional scenario, SR tries to reflect every user's recent interests.
Nevertheless, a KG is generally rather static 
constructed with users' historical information, that is, reflecting long-term interests. Thus, the first challenge is how to better integrate short-term interests modeled from session and long-term ones from the KG for explainable SR. 
(2) Regarding path-guided explainable models, previous studies generally start from the user entity to find reasonable semantic paths in KG. However, besides the user entity (every user might have multiple sessions), SR can also start from any item (interaction-related) in a session. In this case, since the two types of starting point are both particularly complex for SR, how to determine the starting point for each semantic explainable path, is our second challenge. (3) Considering that existing DL-based SR models have already achieved satisfactory performance in terms of recommendation accuracy, our third challenge lies in that, how to assure those DL-based models, equipped with our designed generic framework, can obtain promising results on both recommendation and explanation tasks.

To tackle these issues, we propose a novel Generic \underline{R}einforced \underline{E}xplainable Framework with \underline{K}nowledge Graph for \underline{S}ession-based Recommendation (denoted as REKS). In particular, towards the first challenge, we consider the session information when constructing the KG, and jointly adopt both session and KG information for final session representation.
With regard to the second challenge, we choose to treat the last (interacted) item of every session as the corresponding starting point, because it can better reflect a user's recent interest. Finally, we propose a path-guided reinforcement learning framework with specially designed reward and loss functions to facilitate SOTA, representative non-explainable SR models, where session representations are learned through non-explainable models, 
and then applied to make decisions on determining every next step in identifying semantic paths.

Our contributions are three-fold.
\begin{itemize}
\item As a \emph{generic} framework, REKS has strong generalization ability, which can be easily instantiated with representative non-explainable SR models to well fulfill both the recommendation and explanation tasks.
Specially, REKS adopts a Markov decision process to simultaneously generate recommendation lists and corresponding explainable paths based on KG. 
\item Towards a better framework, technically, REKS considers to start every semantic path from the last item of a session, involves session information in the policy network, and particularly designs suitable
reward and loss functions. Besides, the state vector is also specifically designed by fusing KG and session information.
\item  
REKS can provide explainability for existing non-explainable SR models, meanwhile improve their recommendation performance (accuracy). On the one hand,
exhaustive experiments regarding five SOTA non-explainable SR models (i.e., GRU4REC, NARM, SR-GNN, GCSAN, and BERT4REC) with REKS on four datasets validate the effectiveness of our framework on improving recommendation accuracy. On the other hand,
via a user study and three intuitive examples, we demonstrate that REKS can not only generate reasonable
explanations, but also can better capture real intentions implied by every session.
\end{itemize}

\section{Related Work}
\label{sec:rela}

Our study relates to two areas, namely session-based recommendation and explainable recommendation.

\subsection{Session-based Recommendation}

Early studies on SR focused on exploiting item dependency relationships with Markov chain (MC) \cite{kamehkhosh2017comparison}. For instance, Shani et al. \cite{shani2005mdp} utilized Markov decision process to calculate the transition probability among items. Rendle et al. \cite{rendle2010factorizing} later proposed a hybrid method, i.e., FPMC, which combines matrix factorization and first-order MC to model long-term and short-term user preferences, respectively. Besides the first-order MC, high-order MC is also considered for SR \cite{he2016fusing, he2017translation}.  

In recent years, deep neural network techniques have been applied in SR and achieved encouraging results.
For example, GRU4REC \cite{hidasi2015session} firstly applies RNN (i.e., a multi-layer gate recurrent unit) to process session data.
NARM \cite{li2017neural} explores a hybrid encoder with an attention mechanism to capture the representative item-transition information among items in every session.
Moreover, BERT4REC \cite{sun2019bert4rec} models user behaviors with a bidirectional self-attention network through Cloze task.
However, similar to MC-based methods, RNN- and attention-based methods merely consider item transitions in a single session, and cannot capture item relationship across different sessions.
In this view, GNN-based methods \cite{chen2020handling} have become increasingly popular. For example, SR-GNN \cite{wu2019session} firstly applies
a gated graph network \cite{li2015gated} to learn item embeddings, and adopts a self-attention layer to aggregate item representations. FGNN \cite{qiu2019rethinking} adopts a weighted graph attention network (WGAT) to enhance SR-GNN, whilst GCSAN \cite{xu2019graph} uses both GNN and self-attention mechanism to respectively learn local dependencies and long-range dependencies for SR.

These non-explainable approaches, especially the DL-based ones, have achieved promising performance in terms of recommendation accuracy for session-based recommendation. However, they lack transparency and explanations towards the recommendation results, which thus might lead to lowered user satisfaction and acceptance.

\subsection{Explainable Recommendation}
Recently, explainable recommendation concerning different types of recommender systems (RSs), e.g.,  traditional and session-based RSs, has attracted increasing attention, which mainly consists of two types \cite{zhang2020explainable}: post-hoc (two-stage) \cite{zhang2014explicit} and model-based (end-to-end) methods \cite{lin2018multi,geng2022path}. 
The former one refers to identifying reasonable explanations for the recommended items with auxiliary information after recommendation model ranks items in the first stage, whilst the latter type simultaneously optimizes recommendation and explanation tasks, e.g., in the form of multi-task learning.
For the two types, knowledge graph (KG) has been adopted to facilitate explainable recommendation in traditional scenarios. Both of the two corresponding categories using KG can be concluded as path-based explainable methods, as they obtain path-based explanations from KG for reasonably explaining recommended items.
For example, post-hoc models \cite{ai2018learning,wang2018ripplenet} adopt
KG embedding-based techniques to learn entity/relation representations from built KG to improve recommendation accuracy. Then, explanation paths obtained from KGs are generated to explain recommendations in the second stage.
In the end-to-end models \cite{wang2019explainable}, explanation paths are identified together with recommendation lists, by using neural symbolic reasoning \cite{xian2020cafe} or reinforcement learning \cite{zhao2020leveraging}. 

 Besides, there are a few studies on explainable sequential recommendation\footnote{As have discussed, the major difference between sequential recommendation and session-based recommendation is that the former one considers user information while the latter one generally treats a session as being anonymous.}. For example, Hou et al. \cite{hou2019explainable} proposed a knowledge-aware sequential model that integrates pre-defined path-based and embedding-based methods. However, it is impractical in real-world systems to enumerate all explanatory paths.
Subsequently, RSL-GRU \cite{cui2021reinforced} models the KG-based explainable recommendation in sequential settings with a GRU component and a reinforced path reasoning network (RPRN) component to learn a path search policy. These models are designed for explainable sequential recommendation, and cannot be easily applied to other non-explainable models.
For session-based recommendation, previous explainable studies \cite{chen2021ssr,geng2022causality} are mostly post-hoc methods, where attention mechanisms are harnessed to provide explainability. 

In summary, there is still little research on model-based explainable
SR, howbeit quite a set of non-explainable SR algorithms have achieved SOTA results on recommendation accuracy. Therefore, our research targets to design a generic path-based explainable framework with KG to help existing
non-explainable SR algorithms to achieve explainability and also improve recommendation accuracy.

\begin{table}[t]
\footnotesize
\centering\vspace{-3mm}
\caption{Main notations.}\label{tb:notations}\vspace{-2mm}
\begin{tabular}{@{}c@{}ll}
\toprule
Symbol & Description \\
\midrule
$\mathcal{G}$ \quad  &The knowledge graph.\\ 
$\mathcal{E}$ \quad  &The set of total entities.\\ 
$\mathcal{R}$ \quad  &The set of total relations.\\ 
$\mathcal{U}$ ($\mathcal{V}$) \quad  &The set of users (items), $\mathcal{U} \subset \mathcal{E}$ ($\mathcal{V} \subset \mathcal{E}$).\\
$p(e_i,e_j)$ \quad  &The semantic path from entity $e_i$ to entity $e_j$.\\
$\mathcal{S}_e$ \quad &A session.\\
$\mathbf{X}^0$ \quad &\makecell[l]{The initial representations for all entities and relations.\\$\mathbf{X}^0\in \mathbb{R}^{d_0}$, $d_0$ is the dimension.}\\
$S, A, T, R$ \quad& The state, action, transition, reward function of MDP.\\
$\mathbf{S_e}$ \quad &\makecell[l]{The session representation of $\mathcal{S}_e$ from a non-explainable\\SR model. $\mathbf{S_e}\in \mathbb{R}^{d_1}$, $d_1$ is the dimension.}\\
$\mathbf{S_p}$ \quad 
&\makecell[l]{The path representation from recent entity and relation. \\
$\mathbf{S_p}\in \mathbb{R}^{d_0}$, $d_0$ is the dimension.
}\\
$\pi$  \quad &The probability of selecting next entity.\\
$R_{item}$  \quad & The item-level reward.\\
$R_{rank}$  \quad & The rank-level reward.\\
$R_{path}$  \quad & The path-level reward.\\
$\hat{y}_k$  \quad & The predicted probability of the item $v_k$.\\
$K$ \quad & The length of recommendation list.\\
$L_r$ \quad &The reward loss function.  \\
$L_{ce}$ \quad &The cross-entropy loss function.\\
$P_i$ \quad &The sampling size at step $i$.\\


\bottomrule
\end{tabular}\vspace{-3mm}
\end{table}
\begin{figure*}[htbp]
    \centering
    \includegraphics[width=15.5cm]{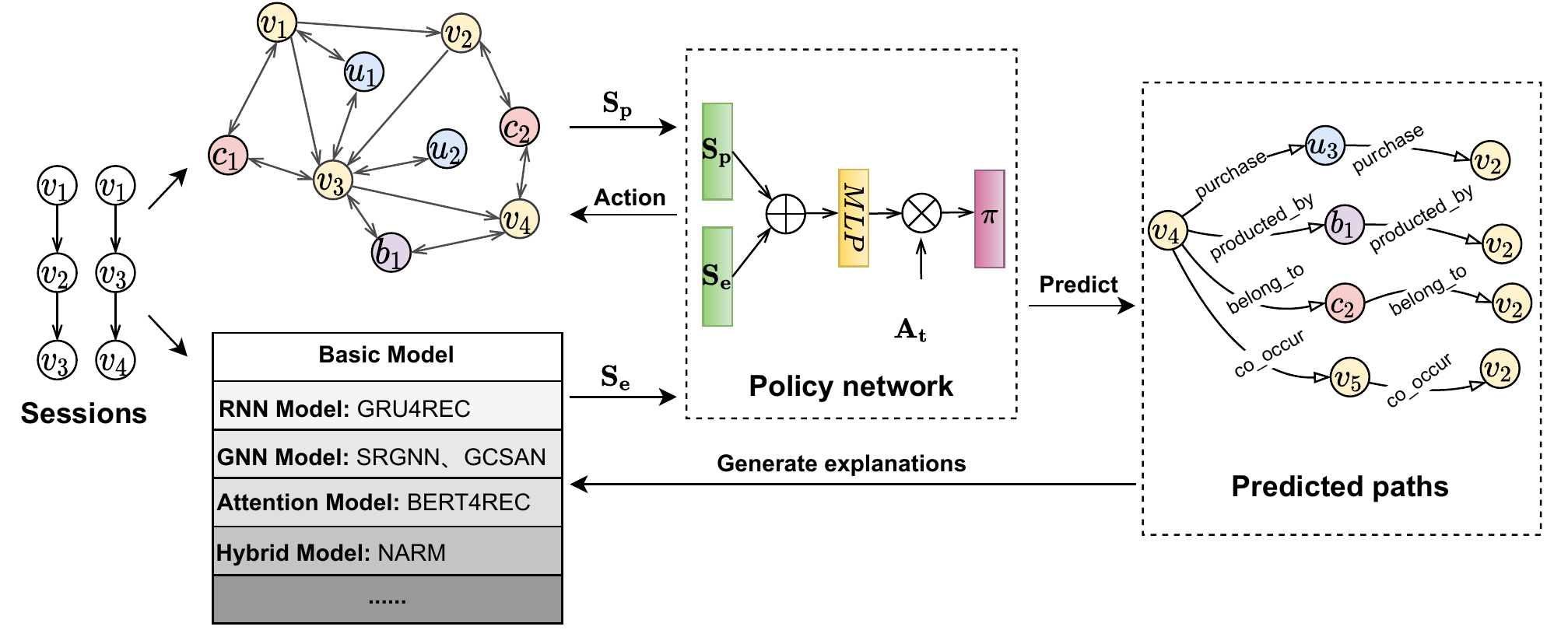}\vspace{-3mm}
    \caption{The overview of our proposed  REKS framework.}\vspace{-2mm}
    \label{fig:model}
\end{figure*}

\section{Our REKS Framework}
\label{sec:model}
In this section, we first formally define some related concepts, including \emph{knowledge graph} and \emph{semantic path}, and the research problem. Then, we present our REKS framework. 

\subsection{Problem Definition}
Before formalizing our research problem, we firstly introduce some concepts related to our work.

\textbf{Knowledge Graph.} A knowledge graph (KG) is formally defined as a graph $\mathcal{G}=\{(e_i,r,e_j)|e_i,e_j \in \mathcal{E}, r \in \mathcal{R}\}$, where $\mathcal{E} = \{e_1, e_2, ..., e_{|\mathcal{E}|}\}$ denotes the set of entities, and $\mathcal{R} = \{r_1, r_2, ..., r_{|\mathcal{R}|}\}$ denotes the set of relations. A triple $(e_i,r,e_j)$ represents a relation $r \in \mathcal{R}$ from a head entity $e_i$ to a tail entity $e_j$ in the KG $\mathcal{G}$. In recommendation scenario, there exist at least two types of entities in $\mathcal{G}$, i.e., the user entity $u \in \mathcal{U} \subset \mathcal{E}$ and the item entity $v \in \mathcal{V} \subset \mathcal{E}$. They are connected by relations like ``purchase'' in e-commerce. Besides, there might be other types of entities and relations (depended on every particular scenario/domain) in $\mathcal{G}$. For example, triple (iPhone, produced\_by, Apple) refers to that iPhone (product entity)'s brand is Apple (brand entity).
  
\textbf{Semantic Path.}
A semantic path between entity $e_i$ and $e_j$ in $\mathcal{G}$ refers to a sequence of entities connected by different types of relations, which is denoted as: $p(e_i,e_j) = {e_i\stackrel{r_{i1}}{\longrightarrow} e_{i+1} \stackrel{r_{i2}}{\longrightarrow} e_{i+2}  \cdots \ \stackrel{r_{ih}}{\longrightarrow} e_j}$, where $e_i\stackrel{r_{i1}}{\longrightarrow} e_{i+1}$ refers to $(e_i,r_{i1},e_{i+1}) \in \mathcal{G}$, and it is meant that $e_i$ and $e_j$ is $h$-hop reachable in $\mathcal{G}$. Noted that there might exist multiple semantic paths between two entities in $\mathcal{G}$ and their lengths are varied.

\textbf{Problem Statement.}
Towards session-based recommendation, let $\mathcal{V}=\{v_1,v_2,\cdots, v_{|\mathcal{V}|}\}$ denote the item set. An anonymous session $\mathcal{S}_e$ can be represented by $\mathcal{S}_e=\{v_1^{S_e},v_2^{S_e},\cdots,v_l^{S_e}\}$, 
where its length equals to $l$ and $v_i^{S_e}\in \mathcal{V}$ means the $i$-th interacted item within $\mathcal{S}_e$. Noted that the items in sessions are the same as the item entities in KG $\mathcal{G}$, which are connected with other entities in terms of different pre-defined (interacted) relations.
Without loss of generality, we define our research problem as follows.
\begin{myDef}
Given the KG $\mathcal{G}=\{(e_i,r,e_j)|e_i,e_j \in \mathcal{E}, r \in \mathcal{R}\}$ and a session $\mathcal{S}_e=\{v_1^{S_e},v_2^{S_e},\cdots,v_l^{S_e}\}$, we aim to predict a probability value for each item $v_k\in \mathcal{V}$, $\widehat{y}_k$.
Correspondingly, the top-$K$ items with the highest values will be recommended as next items that will be most probably interacted given the session $\mathcal{S}_e$ (\textbf{recommendation task}); and for each recommended item, a semantic path is associated to explain why the item is recommended (\textbf{explanation task}). To complete the two tasks, we strive to come up with \emph{a generic reinforced explainable framework}, which is built on existing well-established non-explainable SR models, but provides explainability for these models and also improves their recommendation accuracy. 
\end{myDef}

The main notations are summarized in Table \ref{tb:notations}.


\subsection{The Proposed REKS Model}
Figure \ref{fig:model} outlines the architecture of REKS, which consists of two components: \emph{knowledge graph construction} and \emph{formulating recommendation as a Markov decision process}. In particular, we first build the KG $\mathcal{G}$ and learn the initial entities and relations representation via TransE \cite{bordes2013translating}. Secondly, we formulate our problem as a Markov decision process to simultaneously tackle recommendation and explanation tasks, where we design a policy network to also consider session information besides the KG information, and propose an appropriate reward function to moderately address our tasks.
Next, we elaborate the two components of our model in details.

\begin{table*}[t]
\setlength\tabcolsep{1.8pt}
\footnotesize
\centering \vspace{-3mm}
\caption{Statistics of the relations on the three Amazon datasets.}\label{tb:relationsStatistics}\vspace{-2mm}
\begin{tabular}{lllll}
\toprule
 Relation &Description &Beauty& Cellphones &Baby \\
\midrule

purchase  & $user \xrightarrow{purchase} product$ & 163,982 &137,812  &143,220 \\
produced\_by  & $ product \xrightarrow{produced\_by} brand $&19,356  &10,222  &9,596 \\
belong\_to & $product \xrightarrow{belong\_to} category$  &95,832  &63,178  &13,720 \\
also\_bought &$product \xrightarrow{also\_bought} related\_product$  &1,731,392  &1,127,828  &1,061,056 \\
also\_viewed & $product \xrightarrow{also\_viewed} related\_product$ &1,290,494  &91,138  &723,514 \\
bought\_together &$product \xrightarrow{bought\_together} related\_product$ &28,468  &24,096  &19,110 \\
co\_occur &$product \xrightarrow{co\_occur} product$ &20,199  &16,466  &21,811 \\

\bottomrule
\end{tabular}\vspace{-2mm}
\end{table*}

\begin{table}[t]
\footnotesize\centering\vspace{-3mm}
 \caption{Statistics of the entities on the three Amazon datasets.}\label{tb:entitiesStatistics}\vspace{-2mm}
\begin{tabular}{lllll}
\toprule
Entity& Beauty& Cellphones &Baby \\
\midrule
user \quad&15,438  &17,933  &13,655 \\
product\quad &11,673  &9,805  &6,860 \\
brand \quad &2,008  &904  &716 \\
category \quad&238  &107  &1 \\
related\_product \quad&160,281  &96,674  &68,168 \\

\bottomrule
\end{tabular}\vspace{-2mm}
\end{table}

\subsubsection{Knowledge Graph Construction}
As aforementioned, for each scenario, we firstly construct a KG $\mathcal{G}=\{(e_i,r,e_j)|e_i,e_j \in \mathcal{E}, r \in \mathcal{R}\}$. The entities in Amazon dataset\footnote{In this paper, for better elaboration, we take Amazon dataset as an example to showcase the KG construction process, which can be easily applied in other scenarios/domains. Specifically, we consider three representative and widely-used subsets (domains), i.e., \emph{Beauty}, \emph{Cellphones}, and \emph{Baby}.} can be divided into five types: \emph{user}, \emph{product}, \emph{brand}, \emph{category}, and \emph{related\_product}. 
As shown in Figure \ref{fig:model}, we connect every entity pair with a directed relation if there is a relation from a head entity to a tail entity.
It is worth noting that, in SR, there exists dependent relationship between items (products) in a session. Therefore, 
we accordingly add the \emph{co\_occur} relation in the built KG. That is, with regard to a session $\mathcal{S}_e$, if item $v^{S_e}_i$ is firstly interacted followed by item $v^{S_e}_{i+1}$, we add the triple $(v^{S_e}_i, co\_occur, v^{S_e}_{i+1})$ to $\mathcal{G}$, thus the KG has an edge from $v^{S_e}_i$ to $v^{S_e}_{i+1}$. For other types of relations, similar to \cite{xian2019reinforcement,zhao2020leveraging}, we add a bidirectional edge to the knowledge graph. For example, if there exists a triple $(product, belong\_to, category)$, we add two edges to the $\mathcal{G}$, i.e., $product \xrightarrow{belong\_to}category$ and $category \xrightarrow{belong\_to} product$. 
The statistics of total entities and relations on the three Amazon datasets are summarized in Tables \ref{tb:relationsStatistics}, and \ref{tb:entitiesStatistics}.

After constructing the KG, we leverage Translating Embedding (TransE) technique \cite{bordes2013translating} to obtain the initial representations for all entities and relations. Specifically, we first denote $\mathbf{X}^0 \in \mathbb{R}^{(|\mathcal{E}|+|\mathcal{R}|) \times d_0}$ as initial embedding matrix:
\begin{equation}
\small
\mathbf{X}^0 = \text{TransE}(|\mathcal{E}|+|\mathcal{R}|, d_0)\label{eq:initialnnEmbedding},
\end{equation}
where $d_0$ is the dimension of entity/relation representations, and $|\mathcal{E}|+|\mathcal{R}|$ is the number of total entities and relations.

\subsubsection{Formulating Recommendation as a Markov Decision Process}
We firstly elaborate some preliminaries of our Markov decision process (MDP).
In general, we define the MDP environment with a quadruple $(S, A, T, R)$, where $S$ indicates the state space, $A$ refers to the action space, $T: S \times A \to T$  denotes the state transition function, and $R: S \times A \to R$ indicates the reward function.

\textbf{State.}
Let $s$ ($s \in S$) denote a state in our MDP environment. In our model, the state $s$ consists of two parts, that is, session related information and semantic path related information.

In SR, we can obtain session representation by representative non-explainable SR models, like GRU4REC, NARM, SRGNN, BERT4REC, and GCSAN. For example, using NARM, session representation $\mathbf{S_e}\in\mathbb{R}^{d_1}$ ($d_1$ is the dimension of session representation) is calculated as:
\begin{equation}
\small
\mathbf{S_e}=NARM(\mathbf{X}^0_{\mathcal{V}},\mathcal{S}_e), \label{eq:sessionRepresentation}
\end{equation}
where $\mathbf{X}^0_{\mathcal{V}}$ is the initial embedding of items (measured by TransE in Equation \ref{eq:initialnnEmbedding}), which, together with $\mathcal{S}_e$, contribute to the input of NARM.

Besides, the corresponding semantic path related information (denoted as $\mathbf{S_p}\in\mathbb{R}^{d_0}$) is modeled by the sum of the entity representation $\mathbf{x}^0_{e_t}$ and the relation representation $\mathbf{x}^0_{r_t}$ at time $t$, since the most related information to determine next action in this semantic path is the most recent information (i.e., $e_t$ and $r_t$).

Thus, the representation of the state $s$ at time $t$, $\mathbf{s_t}\in \mathbb{R}^{d_2}$ ($d_2$ is the dimension of the state representation),
is calculated as:
\begin{equation}
\small
\mathbf{s_t}=MLP(\mathbf{S_e} \oplus \mathbf{S_p})\label{eq:pathRepresentation},
\end{equation}
where $\oplus$ denotes concatenation operation, and $MLP(.)$ denotes a multilayer perceptron.

Noted that the initial state representation $\mathbf{s_0}$ at time $t_0$ is represented by session information $\mathbf{S_e}$ and initial entity of the semantic path. Besides, for session-based recommendation, given session $\mathcal{S}_e$, we use the last interacted product, i.e., $v_l^{S_e}$ in the session as the initial entity of the semantic path, considering that the last interacted product is the most recent behavior of this session and thus may involve more information for better next-item prediction \cite{balloccu2022post}. 

\textbf{Action.}
Based on the state $s_t$, we define, at time $t$, an action space $A_t=\{(r_{t+1},e_{t+1})|(e_t,r_{t+1},e_{t+1}) \in  \mathcal{G}\}$, which indicates all possible connections of entity $e_t$ in $\mathcal{G}$ excluding those that have been historically visited in the semantic path.
The action behavior is modeled by a policy network $\pi$, which outputs a probability distribution over actions in $A_t$ to decide the next action (happened at $t+1$). In our REKS model, the policy network is influenced by both session information and semantic path information. Then, we use a softmax function to compute the probability of selecting the next entity: 
\begin{equation}
\small
\pi=softmax(\mathbf{A_t} \odot (\mathbf{W_1} \mathbf{s_t} )),\label{eq:policynetwork}
\end{equation}
where $\mathbf{A_t}$ is a binary vector of $A_t$ and contains the probability of each action, and $\mathbf{W_1}$ is a trainable parameter matrix.

\textbf{Reward.}
It is important to define an appropriate reward for reinforcement learning. For our research, we strive to simultaneously optimize the recommendation and explanation tasks, that is, to accurately recommend top-$K$ products that might be probably preferred/liked, and provide suitable and reasonable explanations regarding recommended items. To fulfill the two objectives, we consider three components in our designed reward function: \emph{item-level reward}, \emph{rank-level reward}, and \emph{path-level reward}:
\begin{equation}
\small
R(s_t,a_t)=R_{item}+2^{R_{rank}}+R_{path}\label{eq:reward},
\end{equation}
where $R_{item}$, $R_{rank}$, and $R_{path}$ denote the item-level reward, rank-level reward, and path-level reward, respectively.

With regard to \emph{item-level reward}, we consider the straightforward reward strategy, namely, the terminal reward, which inclines to give a higher reward if a semantic path ends with the target product (i.e. the ground-truth item that the corresponding user will interact with). 
Here, we define $R_{item}$ with regard to the terminal state $s_T$:
\begin{eqnarray}
\small
R_{item}=\left\{
  \begin{array}{lcl}
 \hspace{-1mm} 1 & &{\hspace{-4mm} if \hspace{-2mm} \quad  e_T= v_{T}}\\
 \hspace{-1mm} \sigma((\mathbf{x}^0_{e_T})^T*\mathbf{x}^0_{v_{T}}) & &{\hspace{-4mm} if \hspace{-2mm} \quad  e_T \neq v_{T} \quad \hspace{-2mm} and \hspace{-2mm} \quad e_T \in \mathcal{V}} \\
 \hspace{-1mm} 0 & &{\hspace{-4mm} otherwise}, \\
  \end{array} \right. \label{eq:itemReward}
\end{eqnarray}
where $v_{T}$ is the ground-truth product,  $\sigma(\cdot)$ is the sigmoid function, and $e_T$ is the last entity on the semantic path, i.e., the predicted entity given the session. The reward function indicates that: if the predicted entity is the target product, $R_{item}=1$; if the predicted entity is not the target product but belongs to the product entity, we set the item-level reward as the similarity between the target product and the predicted entity. Otherwise, we consider the item-level reward to be $0$.

For rank-level reward, we aim to further improve the recommendation accuracy of session-based scenario by pushing the target item to a higher position in the top-K ranking list. Towards this goal, we thus propose our rank-level reward, $R_{rank}$, which is particularly measured as:
\begin{eqnarray}
R_{rank}=\left\{
  \begin{array}{lcl}
   \frac{1}{log(ra_T+2)} & &{if \quad e_T \in \mathcal{V}} \\
  0 & &{otherwise,} \\
  \end{array} \right. \label{eq:rankReward}
\end{eqnarray}
where $ra_T$ refers to the rank of entity $e_T$ in the predicted top-$K$ product list.

Considering path-level reward, we want to generate a semantic path of high quality which is more relevant to the given session and thus can better explain our predicted item at the final step.
Therefore, we define $R_{path}$ as:
\begin{equation}
\small
R_{path}=\sigma(\mathbf{P}^T*\mathbf{S_e}),\label{eq:pathReward}
\end{equation}
where $\sigma(.)$ is the sigmoid function and $\mathbf{P}$ is the representation of this semantic path:
\begin{equation}
\small
\mathbf{P}=mean(\mathbf{x}^0_{e_0},\mathbf{x}^0_{r_1}, \cdots, \mathbf{x}^0_{r_T},\mathbf{x}^0_{e_T}).\label{eq:path}
\end{equation}


\textbf{Transition.}
After obtaining each action $a_t$, the agent can get an intermediate reward, i.e., $R(s_t, a_t)$, which reflects the performance of our model. We then utilize the transition function $T: S \times A \to T$ to update the state. Given the current state $s_t$ and the action $a_t=(r_{t+1},e_{t+1})$, the transition to the next state $s_{t+1}$ is:
\begin{equation}
\small
P[s_{t+1}|s_t,a_t=(r_{t+1},e_{t+1})]=1.\label{eq:transition}
\end{equation}

\textbf{Optimization.}
Given the above MDP environment, we want to learn a stochastic policy $\pi$ that maximizes the expected cumulative reward, and minimizes the cross-entropy loss function. That is, our model's loss function can be made up of two parts: reward loss function (cumulative reward maximization) $L_r$ and cross-entropy loss function $L_{ce}$,
\begin{equation}
\small
L=\beta  L_r +L_{ce} \label{eq:totalLoss},
\end{equation}
where $\beta$ is a hyper-parameter to balance the two parts of loss. The reward  loss function is defined as:
\begin{equation}
L_r=-Q(\theta),
\label{eq:rewardLoss}
\end{equation}
\begin{equation}
\small
Q(\theta)=E_{\pi} \sum_{t=0}^{T-1} \gamma^t R(s_{t+1},a_{t+1}),\label{eq:cumulativeReward}
\end{equation}
where $Q(\theta)$ is the expected cumulative reward, $\theta$ denotes the set of model parameters, and $\gamma$ indicates the discount factor. In this work, we employ REINFORCE with a baseline algorithm \cite{sutton2018reinforcement} to optimize parameters $\theta$.
Then the cross-entropy loss is calculated as:
\begin{equation}
L_{ce}=-\sum ( y_j \log(\hat{y}_j)+(1-y_j)\log (1-\hat{y}_j)),
\label{eq:crossentropyLoss}
\end{equation}
where $y_j$ is the ground-truth score of item $v_j$ ($1$ or $0$). $\hat{y}_j$ is the predicted score, which is generated by the policy network in Equation \ref{eq:policynetwork}.

\begin{algorithm}[t]
\footnotesize
\caption{Overview of the REKS Framework.}\label{alg:algorithm}
\LinesNumbered 
\KwIn{The knowledge graph $\mathcal{G}$ and sessions
}     
\KwOut{Top-$K$ items for each session $\mathcal{S}_e$ and the corresponding explainable paths}
Obtain the initial representations $\mathbf{X}^0$ for all entities and relations via TransE (Equation \ref{eq:initialnnEmbedding})\;
Initialize the model parameters\;
\For{epoch in epochs}
{
    \For{each session $\mathcal{S}_e$}
    {
        \For{t in T (path length)}
        {
            Construct the state at time $t$ and learn its representation $\mathbf{s_t}$, via Equation \ref{eq:pathRepresentation}\;
            Get the probability distribution of entities, $\pi$ (Equation \ref{eq:policynetwork}), and select top-$P_t$ entities\;
            Save the path $p$\;
            Save the reward $R_{item}$, $R_{rank}$, and $R_{path}$;
        }
        Calculate predicted probability $\hat{y}$ for each candidate item and the corresponding path $p$\;
        Compute the loss $L_r$ and $L_{ce}$;
    }
    Implement the Adam optimizer to train the model.
}
\end{algorithm}

After training the model, we implement a probabilistic beam search algorithm \cite{xian2019reinforcement, zhao2020leveraging} to find candidate products and explainable semantic paths given a session in SR. It should be noted that the starting point for the semantic paths is the last interacted product in the session, which is the same as the training process. Algorithm \ref{alg:algorithm} summarizes the pseudocode of the procedure in REKS.

\subsection{Summary and Discussions}
The major contribution of our work is to design \emph{a generic path-based explainable framework} with KG (RKES), which can act as a plugin to provide explanations for existing SR algorithms (e.g., NARM, SR-GNN, GRU4Rec, GCSAN and BERT4Rec), instead of directly proposing a new explainable SR model. As shown in Figure \ref{fig:model}, in our RKES framework, an existing non-explainable SR algorithm (e.g., NARM) helps learn session representation ($\mathbf{S_e}$), which is input to capture state representation in the Markov decision process (MDP), and then applied to make decisions on determining every next step in identifying semantic paths.

Technically, the differences between RKES and previous studies mainly lie in two-fold: (1) in the built KG, we particularly extract ``co\_occur" relationship between items given session information; (2) to further improve the recommendation performance whilst generating reasonable explanations, our RL process combines the short-term interests from the session and long-term ones from the KG, considers to start every semantic path from the last item of a session, and designs advanced reward and loss functions.

\begin{table}[t]
\scriptsize\centering\vspace{-3mm}
 \caption{Statistics of the relations on MovieLens dataset.}\label{tb:relationsStatistics_M}\vspace{-2mm}
\begin{tabular}{lll}
\toprule
 Relation &Description&\#Relations\\
\midrule
belong\_to  \quad&$movie \xrightarrow{belong\_to} genre$  &5,131 \\
directed\_by \quad&$movie \xrightarrow{directed\_by} director$  &2,480 \\
acted\_by  \quad&$movie \xrightarrow{acted\_by} actor$  & 2,090\\
written\_by  \quad&$movie \xrightarrow{written\_by} writer$  &3,393 \\
narrated\_by  \quad&$movie \xrightarrow{narrated\_by} language$  &3,029 \\
rated  \quad&$movie \xrightarrow{rated} rating$  &1,860 \\
produced\_by  \quad&$movie \xrightarrow{produced\_by} country$ &2,212 \\
co\_occur &$movie \xrightarrow{co\_occur} movie$ &20,186 \\

\bottomrule
\end{tabular}\vspace{-2mm}
\end{table}

\begin{table}[t]
\setlength\tabcolsep{1.6pt}
\centering
\footnotesize\vspace{-2mm}
 \caption{Statistics of the entities on MovieLens dataset.}\label{tb:entitiesStatistics_M}\vspace{-2mm}
\begin{tabular}{ccccccccc}
\toprule
entity&movie  & genre & director &actor 
 &writer&language &rating &country\\
\midrule
\#entities&23,475&23	&1,481	&1,196	&2,369	&73  &5  &11\\
\bottomrule
\end{tabular}\vspace{-2mm}
\end{table}
\begin{table}[t]
\centering
\footnotesize\vspace{-3mm}
\caption{Statistics of Amazon and MovieLens datasets.}\label{tb:dataStatistics}\vspace{-2mm}
\begin{tabular}{@{}l@{}llll}
\toprule
Dataset & Beauty& Cellphones &Baby &MovieLens\\
\midrule
\#entities  &189,638  &125,423  &89,400  &7,595\\ 
\#relations &3,349,723  &1,470,740  &1,992,027&40,381\\
\#sessions   &20,830  &24,013  &18,907 &38,016\\ 
\#train sessions   &15,438  &17,933 &13,855 &27,840\\ 
\#validation sessions    &1,898  &2,165  &1,838 &3,801\\ 
\#test sessions   &3,494 &3,915  &3,214 &6,375\\
average length  &3.47  &3.29 &3.88 &3.76\\ 
\bottomrule
\end{tabular}\vspace{-2mm}
\end{table}

\section{Experiments}
\label{sec:experiments}
In this section, we conduct extensive experiments on four datasets to validate the effectiveness of our model, with the goal of answering the four research questions (RQs).
\begin{itemize}
\item \textbf{RQ1}: How does REKS facilitate representative non-explainable SR models on recommendation task?
\item \textbf{RQ2}: How do different components of REKS contribute to the model performance? 
\item \textbf{RQ3}: How do different hyper-parameters affect the performance of REKS?
\item \textbf{RQ4}: How does REKS perform on explanation task?
\end{itemize}
\subsection{Experimental Setup}
\subsubsection{Datasets}
We consider the three aforementioned real-world datasets, i.e., Beauty, Cellphones, and Baby from the Amazon e-commerce platform. Besides, to further validate the effectiveness of RKES, we also consider MovieLens dataset\footnote{\url{grouplens.org/datasets/movielens}.} to conduct comparisons with baselines in terms of recommendation task.
Each dataset consists of both user-item interaction records
and the meta information of users and items. Firstly, we extract different types of entities and relations from metadata. The detailed information are summarized in Tables \ref{tb:relationsStatistics}, \ref{tb:relationsStatistics_M}, \ref{tb:entitiesStatistics} and \ref{tb:entitiesStatistics_M}. For MovieLens dataset, following \cite{wang2019multi}, we use Microsoft Satori to construct the knowledge graph. In particular, we extract movie genre, director, actor, writer, language, rating, and country in total seven categories of information to combine with MovieLens-1M dataset for constructing KG.
Second, similar to \cite{geng2022causality}, we consider a user's interactions that occurred in a day as a session, and filter out items with less than $5$ interactions and sessions with lengths smaller than $2$. Besides, we randomly sample $75\%$ of sessions as the training set, take $10\%$ as the validation set, and the remaining $15\%$ as the test set. Note that, the session information of test set is not involved in each built KG.
The detailed information of the four datasets regarding sessions is reported in Table \ref{tb:dataStatistics}.

\begin{table}[t]
\setlength\tabcolsep{1pt}
\centering
\footnotesize\vspace{-3mm}
\caption{Hyper-parameter setups.}\label{tb:hyperParameter}\vspace{-2mm} 
\begin{tabular}{llcccr}
\toprule
Method  & Dataset & batch size & learning rate &dropout & $\beta$\\
\midrule
\multirow{4}{*}{REKS\_GRU4REC} \quad
&Beauty  \quad &256  &0.001  &0.5 &0.6\\ 
&Cellphones  \quad &32  &0.0001  &0.5 &0.4\\ 
&Baby  \quad &256  &0.0001  &0.7 &0.2\\ 
MovieLens \quad &128  &0.0001  &0.3 &0.2\\
\hline
\multirow{4}{*}{REKS\_NARM} \quad
&Beauty  \quad &256  &0.0005  &0.7 &0.2\\ 
&Cellphones  \quad &32  &0.0001  &0.7 &0.2\\ 
&Baby  \quad &256  &0.0001  &0.7 &0.2\\ 
&MovieLens \quad &32  &0.0001  &0.3  &0.2\\
\hline
\multirow{4}{*}{REKS\_SR-GNN} \quad
&Beauty  \quad &128  &0.001  &0.5 &0.4\\ 
&Cellphones  \quad &256  &0.001  &0.7 &0.6\\ 
&Baby  \quad &256  &0.0001  &0.3 &0.2\\ 
&MovieLens\quad &256  &0.0001  &0.7  &0.4\\
\hline
\multirow{4}{*}{REKS\_GCSAN} \quad
&Beauty  \quad &256  &0.001  &0.5 &0.6\\ 
&Cellphones  \quad &256  &0.005  &0.5 &1.0\\ 
&Baby  \quad &256  &0.0005  &0.7 &0.2\\ 
&MovieLens \quad &256  & 0.005 &0.5 &0.4\\
\hline
\multirow{4}{*}{REKS\_BERT4REC} \quad
&Beauty  \quad &256  &0.0001  &0.7 &0.2\\ 
&Cellphones  \quad &64  &0.0001  &0.7 &0.2\\ 
&Baby  \quad &256  &0.0001  &0.7 &0.2\\ 
&MovieLens \quad &128  &0.001  &0.2  &0.4\\
\bottomrule
\end{tabular}\vspace{-2mm}
\end{table}
\begin{table*}[t]
\setlength\tabcolsep{1pt}
\centering\vspace{-3mm}
\caption{Overall performance comparison on the four datasets.
}\label{tb:comparativeResults}\vspace{-2mm}
\resizebox{\textwidth}{35mm}{
\begin{tabular}{c|c|ccr|ccr|ccr|ccr|ccr }
\hline

 Dataset &Metric & GRU4REC & \makecell[c]{REKS\_\\GRU4REC} & Improv.  & NARM  & \makecell[c]{REKS\_\\NARM} & Improv.& SR-GNN & \makecell[c]{REKS\_\\SR-GNN} & Improv.  & GCSAN  & \makecell[c]{REKS\_\\GCSAN} & Improv.& BERT4REC  & \makecell[c]{REKS\_\\BERT4REC} & Improv. \\
\hline

\multirow{6}{*}{Beauty}
&  HR@5 &8.70  &9.91  &13.91\%**  &9.48  &10.01  &5.59\%**  &9.43  &10.01  &6.15\%**  &7.87  &10.20  &29.61\%**  &9.40  &9.99  &6.28\%** \\
&  HR@10 &10.98  &13.16  &19.85\%**  &11.88  &13.92  &17.17\%**  &11.62  &13.47  &15.92\%**  &10.49  &13.33  &27.07\%**  &12.41  &13.69  &10.31\%** \\
&  HR@20 &14.00  &17.52  &25.14\%**  &14.67  &17.85  &21.68\%**  &14.21  &17.64  &24.14\%**  &13.09  &17.27  &31.93\%**  &15.88  &18.17  &14.42\%** \\
& NDCG@5 &6.49   &6.71  &3.39\%**  &7.12  &7.38  &3.65\%**  &6.92  &7.00  &1.16\%	
&5.76  &7.00  &21.53\%**  &6.78  &6.98  &2.95\%**\\
& NDCG@10 &7.23  &7.92  &9.54\%**  &7.89  &8.51   &7.86\%**  &7.63  &8.12  &6.42\%**
&6.61  &8.01  &21.18\%**  &7.77  &8.18  &5.28\%** \\
& NDCG@20 &7.99  &9.04  &13.14\%**  &8.59  &9.50  &10.59\%**  &8.28 &9.16  &10.63\%**
&7.26  &9.01  &24.10\%**  &8.64  &9.30  &7.64\%** \\
\hline
\multirow{6}{*}{Cellphones}
&  HR@5  &7.22 &8.04 &11.36\%** &8.26 &8.52   &3.15\%** &7.16 &7.42 &3.63\%** &6.85  &7.36 &7.45\%**   &7.63  &7.88   &3.28\%**       \\
&  HR@10  &9.62  &11.20 &16.42\%**  &10.98 &11.71  &6.65\%** &9.87 &10.23 &3.65\%** &9.48  & 10.45 &10.23\%**  &10.91 &11.47 &5.13\%**     \\
&  HR@20  &12.91 &15.12  &17.12\%**  &14.30 &15.59 &9.02\%** &13.23 &14.18 &7.18\%** &12.81 &14.63  &14.21\%** &14.47 &15.78 &9.05\%** \\
& NDCG@5   &5.10  &5.34 &4.71\%**  &5.78 &5.85  &1.21\% &5.06  &5.18 &2.37\%**    &4.66 &5.16  &10.73\%** &5.34  &5.40  &1.12\% \\
& NDCG@10   &5.87   &6.34 &8.01\%** &6.71  &6.88  &2.53\%** &5.95 &6.09 &2.35\%**  &5.50 &6.15  &11.82\%**  &6.36 &6.55 &2.99\%** \\
& NDCG@20   &6.70  &7.33  &9.40\%**  &7.55  &7.86 &4.11\%** &6.79 &7.08 &4.27\%**   &6.34 &7.20  &13.56\%** &7.27 &7.64 &5.09\%**  \\
\hline

\multirow{6}{*}{Baby}
&  HR@5  &4.83  &5.32  &10.14\%** &5.15  &5.42  &5.24\%**  &3.90 &5.24 &34.36\%**  &3.56  &5.14 &44.38\%**  &5.12 &5.24 &2.34\%**  \\
&  HR@10  &7.41  &7.96 &7.42\%**  &7.37   &7.70  &4.48\%**  &6.22  &7.63  &22.67\%** &5.97 &6.94 &16.25\%**   &7.47  &7.84  &4.95\%**  \\
&  HR@20  &10.92 &11.05  &1.19\%* &11.01  &11.04  &0.27\%  &9.31  &10.79  &15.90\%**  &9.63  &10.15 &5.40\%** &10.68  &11.03  &3.28\%** \\
& NDCG@5 &2.86  &3.68  &28.67\%**  &3.25 &3.75  &15.38\%**  &2.44  &3.56  &45.90\%**  &2.43 &3.55  &46.09\%** &3.18   &3.63  &14.15\%** \\
& NDCG@10 &3.69  &4.53 &22.76\%**  &3.96  &4.49 &13.38\%** &3.18  &4.33 &36.16\%** &3.01  &4.10 &36.21\%**  &3.93  &4.45 &13.23\%**\\
& NDCG@20  &4.57 &5.31  &16.19\%**  &4.89 &5.33  &9.00\%**  &3.96 &5.14 &29.80\%**  &3.92 &4.90 &25.00\%**  &4.72  &5.25  &11.23\%**  \\
\hline


\multirow{6}{*}{MovieLens}
&  HR@5  &5.89  &6.15 &4.41\%** &6.00  &6.05  &0.83\%   &5.70 &6.03 &5.79\%**  &5.45  &5.50 &0.92\%  &5.73 &5.83 &1.75\%*  \\
&  HR@10   &11.21 &12.62 &12.58\%** &11.91 &12.88 &8.14\%** &10.02 &12.01 &19.89\%**  &10.28  &10.66 &3.70\%**  &11.09& 12.38&11.63\%**  \\
&  HR@20   &18.91 &22.96 &21.42\%**&19.15 &24.79 &29.45\%**  &16.23 &24.30 &49.72\%**  &16.08  &23.92 &48.76\%**  &18.80&21.20 &12.77\%**  \\
& NDCG@5 &3.66 &4.02 &9.84\%**  &3.99 &4.13 &3.51\%**&3.57 &3.75 &5.04\%**  &3.36 &3.54 &5.36\%**  &3.72&3.87 &4.03\%**  \\
& NDCG@10  &5.38 &6.50 &20.82\%** &5.85 &6.79 &16.07\%**  &5.15 &6.46 &25.44\%**  &4.90  &5.67 &15.71\%**  &5.43&6.44 &18.60\%**  \\
& NDCG@20    &8.93 &11.44 &28.11\%** &9.31 &11.94 &28.25\%**  &6.70 &11.07 &65.22\%**&6.35  &10.71 &68.66\%**    &8.38&11.19 &33.53\%**  \\
\hline
\end{tabular}}\vspace{-2mm}
\end{table*}

\input{table/Loss}
\begin{figure*}[tb]
    \centering
	\footnotesize
	\begin{tikzpicture}
      \matrix[
          matrix of nodes,
          draw,
          inner sep=0.2em,
          ampersand replacement=\&,
          font=\tiny,
          anchor=south
        ]
        { 
		\ref{plots:REKS_GRU-user} REKS\_GRU4REC\_user
		\ref{plots:REKS_GRU1} REKS\_GRU4REC
		\ref{plots:REKS_NARM-user} REKS\_NARM\_user
		\ref{plots:REKS_NARM1} REKS\_NARM
		\ref{plots:REKS_SR-GNN-user} REKS\_SR-GNN\_user\\
		\ref{plots:REKS_SR-GNN1} REKS\_SR-GNN 
		\ref{plots:REKS_GCSAN-user} REKS\_GCSAN\_user
		\ref{plots:REKS_GCSAN1} REKS\_GCSAN
		\ref{plots:REKS_BERT-user} REKS\_BERT4REC\_user
		\ref{plots:REKS_BERT1} REKS\_BERT4REC\\
          };
    \end{tikzpicture}\\

	\begin{tikzpicture}
	\begin{groupplot}[group style={
		group name=myplot,
		group size= 2 by 2,  horizontal sep=0.9cm,vertical sep=0.7cm}, 
		height=3.4cm, width=9.3cm,
	ylabel style={yshift=-0.15cm},
	tick align=outside,
	every tick label/.append style={font=\tiny}
	]

	\nextgroupplot[ybar=0.10,
	bar width=0.38em,
	ylabel={HR@5},
	scaled ticks=false,
	ymin=0, ymax=12,
	enlarge x limits=0.3,
	symbolic x coords={Beauty,Cellphones,Baby},
	ylabel style = {font=\scriptsize},
    xlabel style = {font=\scriptsize},
	legend style={font=\scriptsize},
	xtick=data,
	ytick={0,5,10},
	]
	\addplot coordinates {
		(Beauty,3.99)   (Cellphones,2.44) (Baby,1.67)
		}; \label{plots:REKS_GRU-user}
	\addplot  coordinates {
		(Beauty,9.91)  (Cellphones,8.04) (Baby,5.32)
		}; \label{plots:REKS_GRU1}
	\addplot   coordinates {
		(Beauty,0)  (Cellphones,0) (Baby,0)
		};  
	\addplot [fill=lightkhaki]   coordinates {
		(Beauty,4.23)  (Cellphones,2.87) (Baby,1.80)
		};   \label{plots:REKS_NARM-user}
	\addplot   coordinates {
		(Beauty,10.01)  (Cellphones,8.52) (Baby,5.42)
		}; \label{plots:REKS_NARM1}
	\addplot   coordinates {
		(Beauty,0)  (Cellphones,0) (Baby,0)
		}; 
		\addplot [fill=orange(colorwheel)]   coordinates {
		(Beauty,3.49)  (Cellphones,2.56) (Baby,1.36)
		};  \label{plots:REKS_SR-GNN-user}
        \addplot [fill=oldlavender]  coordinates {	
		(Beauty,10.01)  (Cellphones,7.42) (Baby,5.24)
		}; \label{plots:REKS_SR-GNN1}
	\addplot   coordinates {
		(Beauty,0)  (Cellphones,0) (Baby,0)
		}; 
	\addplot [fill=shockingpink]  coordinates {
	
		(Beauty,2.35)  (Cellphones,2.67) (Baby,1.24)}; \label{plots:REKS_GCSAN-user}
	\addplot [fill=lavender(floral)] coordinates {
	
		(Beauty,10.20)  (Cellphones,7.36) (Baby,5.14)
		}; \label{plots:REKS_GCSAN1}
	\addplot  coordinates {
		(Beauty,0)  (Cellphones,0) (Baby,0)
		}; 
	\addplot  [fill=oldgold] coordinates {
		(Beauty,4.29)  (Cellphones,2.78) (Baby,1.42)
		};  \label{plots:REKS_BERT-user}
	\addplot [fill=thistle]  coordinates {
		(Beauty,9.99)  (Cellphones,7.88) (Baby,5.24)
		};  \label{plots:REKS_BERT1}
		

  	\nextgroupplot[ybar=0.10,
	bar width=0.38em,
	ylabel={NDCG@5},
	scaled ticks=false,
	ymin=0, ymax=11,
	enlarge x limits=0.3,
	symbolic x coords={Beauty,Cellphones,Baby},
	ylabel style = {font=\scriptsize},
    xlabel style = {font=\scriptsize},
	legend style={font=\scriptsize},
	legend pos=outer north east, 
	xtick=data,
	ytick={0,3,6,9},
	]

	\addplot coordinates {
		(Beauty,2.64)  (Cellphones,1.65) (Baby,0.98)
		};
	\addplot  coordinates {
		(Beauty,6.71)  (Cellphones,5.34) (Baby,3.68)
		};
	\addplot   coordinates {
		(Beauty,0)  (Cellphones,0) (Baby,0)
		}; 
	\addplot [fill=lightkhaki]  coordinates {
	(Beauty,2.91)  (Cellphones,1.84) (Baby,1.09)
		};  
	\addplot   coordinates {
		(Beauty,7.38)  (Cellphones,5.85) (Baby,3.75)
		};
	\addplot   coordinates {
		(Beauty,0)  (Cellphones,0) (Baby,0)
		}; 
	\addplot   [fill=uclagold]  coordinates {
		(Beauty,2.40)  (Cellphones,1.73) (Baby,0.87)
		};  
	\addplot [fill=oldlavender]  coordinates {
		(Beauty,7.00)  (Cellphones,5.18) (Baby,3.56)
		};
	\addplot   coordinates {
		(Beauty,0)  (Cellphones,0) (Baby,0)
		}; 

		\addplot [fill=shockingpink]  coordinates {
        (Beauty,1.56)  (Cellphones,1.73) (Baby,0.78)
		};  
	\addplot [fill=lavender(floral)] coordinates {
		(Beauty,7.00)  (Cellphones,5.16) (Baby,3.55)
		};
	\addplot   coordinates {
		(Beauty,0)  (Cellphones,0) (Baby,0)
		}; 
	\addplot [fill=oldgold] coordinates {
		(Beauty,2.87)  (Cellphones,1.75) (Baby,0.93)
		};  
	\addplot [fill=thistle]  coordinates {
		(Beauty,6.98)  (Cellphones,5.40) (Baby,3.63)
		};
	\nextgroupplot[ybar=0.10,
	bar width=0.38em,
	ylabel={HR@10},
	scaled ticks=false,
	ymin=0, ymax=17,
	enlarge x limits=0.3,
	symbolic x coords={Beauty,Cellphones,Baby},
	ylabel style = {font=\scriptsize},
    xlabel style = {font=\scriptsize},
	xtick=data,
	ytick={0,5,10,15},
	]
	\addplot coordinates {
		(Beauty,5.97)   (Cellphones,4.09) (Baby,2.77)
		}; 
	\addplot  coordinates {
		(Beauty,13.16)  (Cellphones,11.20) (Baby,7.96)
		}; 
	\addplot   coordinates {
		(Beauty,0)  (Cellphones,0) (Baby,0)
		}; 
	\addplot  [fill=lightkhaki]  coordinates {
		(Beauty,6.73)  (Cellphones,4.27) (Baby,2.79)
		};   
	\addplot   coordinates {
		(Beauty,13.92)  (Cellphones,11.71) (Baby,7.70)
		}; 
	\addplot   coordinates {
		(Beauty,0)  (Cellphones,0) (Baby,0)
		}; 
		\addplot [fill=orange(colorwheel)]   coordinates {
		(Beauty,5.67)  (Cellphones,4.01) (Baby,2.13)
		};  
        \addplot [fill=oldlavender]  coordinates {	
		(Beauty,13.47)  (Cellphones,10.23) (Baby,7.63)
		}; 
	\addplot   coordinates {
		(Beauty,0)  (Cellphones,0) (Baby,0)
		}; 
	\addplot [fill=shockingpink]  coordinates {
	
		(Beauty,4.43)  (Cellphones,4.24) (Baby,2.09)}; 
	\addplot [fill=lavender(floral)] coordinates {
	
		(Beauty,13.33)  (Cellphones,10.45) (Baby,6.94)
		}; 
	\addplot  coordinates {
		(Beauty,0)  (Cellphones,0) (Baby,0)
		}; 
	\addplot  [fill=oldgold] coordinates {
		(Beauty,6.67)  (Cellphones,4.25) (Baby,2.38)
		};  
	\addplot [fill=thistle]  coordinates {
		(Beauty,13.69)  (Cellphones,11.47) (Baby,7.84)
		};  
		

  	\nextgroupplot[ybar=0.10,
	bar width=0.38em,
	ylabel={NDCG@10},
	scaled ticks=false,
	ymin=0, ymax=11,
	enlarge x limits=0.3,
	symbolic x coords={Beauty,Cellphones,Baby},
	ylabel style = {font=\scriptsize},
    xlabel style = {font=\scriptsize},
	legend style={font=\scriptsize},
	legend pos=outer north east, 
	xtick=data,
	ytick={0,3,6,9},
	]

	\addplot coordinates {
		(Beauty,3.27)  (Cellphones,2.18) (Baby,1.34)
		};
	\addplot  coordinates {
		(Beauty,7.92)  (Cellphones,6.34) (Baby,4.53)
		};
	\addplot   coordinates {
		(Beauty,0)  (Cellphones,0) (Baby,0)
		}; 
	\addplot [fill=lightkhaki]  coordinates {
	(Beauty,3.71)  (Cellphones,2.29) (Baby,1.41)
		};  
	\addplot   coordinates {
		(Beauty,8.51)  (Cellphones,6.88) (Baby,4.49)
		};
	\addplot   coordinates {
		(Beauty,0)  (Cellphones,0) (Baby,0)
		}; 
	\addplot   [fill=uclagold]  coordinates {
		(Beauty,3.10)  (Cellphones,2.19) (Baby,1.12)
		};  
	\addplot [fill=oldlavender]  coordinates {
		(Beauty,8.12)  (Cellphones,6.09) (Baby,4.33)
		};
	\addplot   coordinates {
		(Beauty,0)  (Cellphones,0) (Baby,0)
		}; 

		\addplot [fill=shockingpink]  coordinates {
        (Beauty,2.24)  (Cellphones,2.23) (Baby,1.05)
		};  
	\addplot [fill=lavender(floral)] coordinates {
		(Beauty,8.01)  (Cellphones,6.15) (Baby,4.10)
		};
	\addplot   coordinates {
		(Beauty,0)  (Cellphones,0) (Baby,0)
		}; 
	\addplot [fill=oldgold] coordinates {
		(Beauty,3.63)  (Cellphones,2.22) (Baby,1.25)
		};  
	\addplot [fill=thistle]  coordinates {
		(Beauty,8.18)  (Cellphones,6.55) (Baby,4.45)
		};

	\end{groupplot}
    \end{tikzpicture}\vspace{-3mm}
    \caption{Impact of starting point ($K=5,10$).}
    \label{fig:startingpoint}\vspace{-2mm}
\end{figure*}
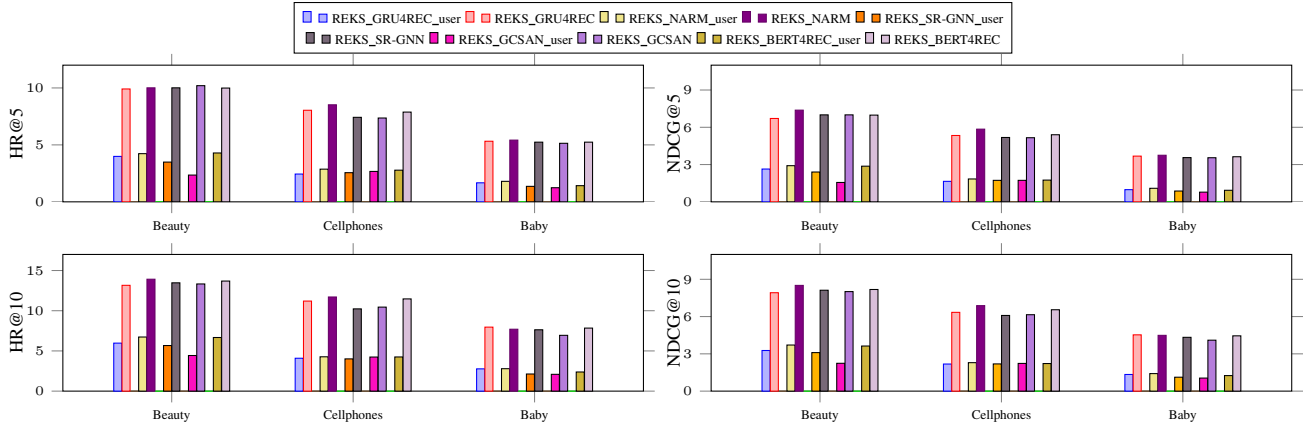
\subsubsection{Baseline Models}
We apply our REKS model on five representative and widely-used SR baselines, i.e., one RNN-based method (\textbf{GRU4REC}), two GNN-based models (\textbf{SR-GNN} and \textbf{GCSAN}), one attention-based method (\textbf{BERT4REC}), and one hybrid method that combines attention and RNN (\textbf{NARM}):
\textbf{GRU4REC} \cite{hidasi2015session} applies GRU to model interaction sequences and designs a ranking-based loss function; \textbf{SR-GNN} \cite{wu2019session} is a solid GNN-based approach for SR, which employs a gated graph convolutional layer to obtain item embedding and applies a self-attention mechanism to obtain session representation; \textbf{GCSAN} \cite{xu2019graph} utilizes self-attention mechanism to capture item dependencies via graph information aggregation; \textbf{BERT4REC} \cite{sun2019bert4rec} employs a deep bidirectional self-attention framework to model user behaviors and utilizes a Cloze objective loss for SR; \textbf{NARM} \cite{li2017neural} improves GRU4REC by utilizing vanilla attention to model the relationship of the last item in the session with other items of the session.

We compare each of the five baselines with the corresponding enhanced version with REKS, respectively (e.g., NARM vs. RKES\_NARM).
Noted that for fair comparisons, the inputs of these baselines are the same as the corresponding enhanced version with REKS, including both the knowledge graph $\mathcal{G}$ and session information. We also apply TransE method on $\mathcal{G}$ to obtain the initial item representations for these baselines without REKS.

\subsubsection{Evaluation Metrics}
Given a session, each method generates a top-$K$ recommendation list ($K$ is set to $5$, $10$, $20$ in our experiments).
Towards exhaustive evaluations on the \emph{recommendation task}, we consider two widely used accuracy metrics: \textbf{HR}@$K$ and \textbf{NDCG}@$K$. \textbf{HR}@$K$ indicates the hit ratio, i.e., the coverage rate of targeted predictions; 
and \textbf{NDCG}@$K$ (Normalized Discounted Cumulative Gain) rewards each hit based on its position in the ranked recommendation list. A larger value indicates better performance for the two metrics.
Regarding the \emph{explanation task}, we firstly conduct a questionnaire-based survey to evaluate user satisfaction towards the explanations generated by our framework. Then, we also present three explanation cases to intuitively elaborate on how REKS interprets
the recommendation results.

\subsubsection{Hyper-parameter Setups}
Regarding each method, we tune the best hyper-parameters according to validation sessions on each dataset. For all methods, dimension $d_0$, $d_1$, and $d_2$ are set to $400$ on three Amazon datasets, and $64$ on MovieLens dataset. 
We train all models with Adam optimizer. The path length is $2$ and the sampling size of each step is set to $100$ and $1$, respectively on each dataset. 
The discount factor is $0.99$. The settings of other parameters are shown in Table \ref{tb:hyperParameter}.
Noted that for all these methods, we run each experiment five times and report the average as the performance in Table \ref{tb:comparativeResults}, where REKS\_baseline denotes applying RKES with the corresponding baseline. We also conduct a paired t-test (regarding every baseline and REKS\_baseline pair)
to validate the significance of the performance difference ($^{*}$ for p-value $\leq$.05 and $^{**}$ for p-value $\leq$.01). Besides, the results are reported in percentage (\%).

\input{table/Reward}
\input{table/Length}
\subsection{Experimental Results on Recommendation Task}

Here, we present results to answer RQ1, RQ2 and RQ3.
\subsubsection{Performance of REKS with Baselines (RQ1)}
We instantiate the REKS framework on five SOTA non-explainable SR methods, and their performance on four real-world datasets is reported in Table \ref{tb:comparativeResults}.
From Table \ref{tb:comparativeResults}, we have some interesting observations as below.

(a) With REKS framework, the performance of every baseline can be significantly improved in almost all cases, validating the effectiveness of our framework for non-explainable SR methods on recommendation task. For example, on Beauty, REKS\_GRU4REC achieves 9.91\% in terms of HR@5 with an improvement of 13.91\% over vanilla GRU4REC (8.70\%).
(b) Among non-explainable SR models, NARM performs better than other methods on Cellphones, Baby and MovieLens, while BERT4REC performs the best on Beauty.
Besides, surprisingly, vanilla GCSAN obtains the worst 
performance across all scenarios. However, in most cases, RKES\_GCSAN also makes the largest progress over GCSAN compared to other methods, indicating that our RL-based framework can well mitigate the performance gaps among non-explainable SR models.
(c) REKS\_baseline inclines to get better performance than other REKS equipped methods if the corresponding baseline performs better, implying that a better session representation learned from non-explainable SR methods can better guide our RL-based framework to find much more reasonable paths for recommendation task in SR.

To conclude, our REKS framework can significantly improve the performance of non-explainable SR methods on recommendation task.


\subsubsection{Ablation Study (RQ2)}
We conduct experiments to test the innovative designs in REKS: (1) reward function; (2) loss function; (3) starting point of semantic paths; (4) path length; and (5) user information in the KGs.

\textbf{Impact of reward function.}
In REKS, we innovatively design a complex reward function for the recommendation and explanation tasks. It consists of three parts: item-level, path-level, and rank-level rewards. Thus, we compare our model with three variants: (1) REKS-rank does not consider rank-level reward; (2) REKS-path further removes the path-level reward from REKS-rank, i.e., only uses item-level reward; and (3) REKS\_R1 merely employs the most basic reward, which equals $1$ if the predicted product is the ground-truth product. Figure \ref{fig:rewardfunction} 
shows their performance
, where the x-label represents the corresponding non-explainable SR model. As we can see, each component contributes to the final performance, validating the effectiveness of our reward designs for making more accurate recommendation.    

\textbf{Impact of loss function.}
The loss function in REKS consists of two parts: reward loss and cross entropy loss. To validate the effectiveness of each part, we compare our model with two alternatives: REKS\_R merely considers reward loss, while REKS\_C only uses cross entropy loss. The results with different loss functions are shown in Figure \ref{tb:differentlossfunction}. REKS\_R and REKS\_C perform consistently worse than REKS, which demonstrates that each loss function can improve recommendation accuracy. Besides, REKS\_R performs better than REKS\_C, which might be caused by that, cross entropy loss function is directly linked to recommendation accuracy for recommendation task, whereas in reward loss (reward maximization), we further consider to address explanation task (e.g., path-level reward design).

\textbf{Impact of starting point of semantic paths.}
As discussed in Section \ref{sec:intro}, the starting point of semantic paths can be a user or a product (the last interacted item of a session), while in our REKS, we choose to start from the last interacted product given a session. Hence, to verify the validity of our design, we compare our framework (REKS) with a variant, REKS\_user, starting from the user of the session. Noted that for each REKS\_user model, we re-tune the corresponding best hyper-parameters: path length is $3$, and the sampling size is set to $\{100,10,1\}$ for each step. 
From the results shown in Figure \ref{fig:startingpoint}, we observe that REKS consistently performs better than REKS\_user, validating our hypothesis that, for SR, the last interacted product of a session can better reflect the corresponding user's recent interests, and thus lead to more accurate next-item prediction.

\textbf{Impact of path length.}
Here, we check the effect of path length on the model performance, while in our REKS, we prefer path length of $2$. We thus compare our model with two variants: REKS with path length being $3$ (REKS\_l3) and that of length being $4$ (REKS\_l4), where the sampling size is $\{100,1,1\}$, and $\{100,1,1,1\}$ for each step, respectively.
Figure \ref{fig:differentpathlength} shows the comparative results, and we can see that when path length is $2$, REKS can obtain the best recommendation accuracy. This might be caused that, though a larger path length can generate more patterns, it may also introduce noisy information. Besides, REKS\_l4 performs better than REKS\_l3 in most cases. This is because a path pattern is mostly ``product $\to$ other entity $\to$ product $\to$ other entity'' in the knowledge graph, we are thus expected to get better recommendation performance when the path length is an even number.
\begin{figure*}[tbp]
    \centering
	\footnotesize
	\begin{tikzpicture}
      \matrix[
          matrix of nodes,
          draw,
          inner sep=0.2em,
          ampersand replacement=\&,
          font=\tiny,
          anchor=south
        ]
        { 
		\ref{plots:Beauty} Beauty
		\ref{plots:Cellphones} Cellphones
		\ref{plots:Baby} Baby\\
          };
    \end{tikzpicture}\\
\begin{tikzpicture}
	\begin{groupplot}[group style={
		group name=myplot,
		group size= 4 by 1,  horizontal sep=0.9cm}, 
		height=3.4cm, width=5cm,
	ylabel style={yshift=-0.15cm},
	tick align=outside,
	every tick label/.append style={font=\tiny}
	]
	
\nextgroupplot[
ylabel=HR@10,
xlabel=lr,
ymax=15.5,
ytick={3,6,9,12,15},
scaled ticks=false,
symbolic x coords={1e-4,5e-4,1e-3,5e-3},
xtick=data,
line width=1pt,
ylabel style = {font=\scriptsize},
xlabel style = {font=\scriptsize},
ytick pos=left
]
\addplot[color=blue,mark=x] coordinates {
(1e-4,13.70)
(5e-4,13.92)
(1e-3,13.25)
(5e-3,13.53)
};\label{plots:Beauty}
\addplot[color=orange,mark=square] coordinates {
(1e-4,11.71)
(5e-4,11.36)
(1e-3,10.93)
(5e-3,10.91)
};\label{plots:Cellphones}
\addplot[color=olive,mark=o] coordinates {
(1e-4,7.70)
(5e-4,7.32)
(1e-3,6.97)
(5e-3,6.38)
};\label{plots:Baby}

\nextgroupplot[
ylabel=NDCG@10,
xlabel=lr,
ymax=10.5,
ytick={2,4,6,8,10},
scaled ticks=false,
symbolic x coords={1e-4,5e-4,1e-3,5e-3},
xtick=data,
line width=1pt,
ylabel style = {font=\scriptsize},
xlabel style = {font=\scriptsize},
ytick pos=left
]
\addplot[color=blue,mark=x] coordinates {
(1e-4,8.20)
(5e-4,8.51)
(1e-3,8.125)
(5e-3,8.18)
};
\addplot[color=orange,mark=square] coordinates {
(1e-4,6.88)
(5e-4,6.76)
(1e-3,6.53)
(5e-3,6.21)
};
\addplot[color=olive,mark=o] coordinates {
(1e-4,4.49)
(5e-4,4.30)
(1e-3,4.14)
(5e-3,3.69)
};

\nextgroupplot[
ylabel=HR@10,
xlabel=$\beta$,
ymax=15.5,
ytick={6,9,12,15},
scaled ticks=false,
symbolic x coords={0.2,0.4,0.6,0.8,1.0,1.2},
xtick=data,
ylabel style = {font=\scriptsize},
xlabel style = {font=\scriptsize},
line width=1pt,
ytick pos=left
]
\addplot[color=blue,mark=x] coordinates {
(0.2,13.92)
(0.4,13.44)
(0.6,13.74)
(0.8,13.27)
(1.0,13.09)
(1.2,12.83)
};
\addplot[color=orange,mark=square] coordinates {
(0.2,11.71)
(0.4,11.61)
(0.6,11.39)
(0.8,11.31)
(1.0,11.01)
(1.2,11.12)
};
\addplot[color=olive,mark=o] coordinates {
(0.2,7.70)
(0.4,7.20)
(0.6,7.18)
(0.8,6.73)
(1.0,7.00)
(1.2,7.09)
};

\nextgroupplot[
ylabel=NDCG@10,
xlabel=$\beta$,
ymax=10.5,
ytick={4,6,8,10},
scaled ticks=false,
symbolic x coords={0.2,0.4,0.6,0.8,1.0,1.2},
xtick=data,
ylabel style = {font=\scriptsize},
xlabel style = {font=\scriptsize},
line width=1pt,
ytick pos=left
]
\addplot[color=blue,mark=x] coordinates {
(0.2,8.51)
(0.4,8.27)
(0.6,8.40)
(0.8,8.16)
(1.0,8.01)
(1.2,7.91)
};
\addplot[color=orange,mark=square] coordinates {
(0.2,6.88)
(0.4,6.86)
(0.6,6.77)
(0.8,6.65)
(1.0,6.59)
(1.2,6.43)
};
\addplot[color=olive,mark=o] coordinates {
(0.2,4.49)
(0.4,4.28)
(0.6,4.33)
(0.8,3.96)
(1.0,3.95)
(1.2,3.89)
};

\end{groupplot}

\end{tikzpicture}\vspace{-3mm}
  \caption{Impact of different hyper-parameters on REKS\_NARM model ($K$=10).}
\label{fig:hyper}\vspace{-2mm}
\end{figure*}
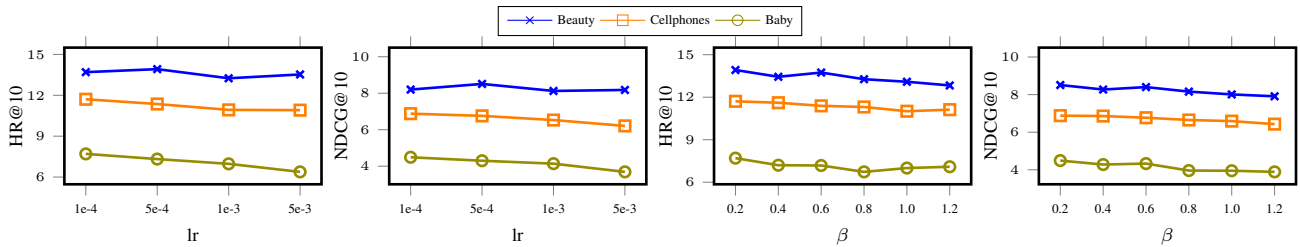
\begin{table}[t]
\setlength\tabcolsep{1.2pt}
\centering\vspace{-2mm}
 \caption{Comparative results on the KGs without user information.} \label{tb:userInformation}\vspace{-2mm}
\begin{tabular}{c|c|ccc|ccc}
\hline
 & &\multicolumn{3}{c|}{HR}&\multicolumn{3}{c}{NDCG}\\
\cline{3-8}
\hline
 Dataset &Method & @5 & @10 & @20 & @5 & @10 & @20 \\

\hline
\multirow{2}{*}{Beauty}
& REKS &9.11	&11.38	&14.06	&6.88	&7.63	&8.30  \\
&REKS\_NARM &9.24	&12.20	&16.24	&6.96	&7.92	&8.73   \\

\hline
\multirow{2}{*}{Cellphones}
& REKS&8.02	&10.90	&13.57	&5.01	&6.27	&7.44\\
&REKS\_NARM &8.12	&11.06	&14.55	&5.40	&6.62	&7.57\\

\hline

\multirow{2}{*}{Baby}
& REKS &4.54	&6.62	&9.40	&3.11	&3.78	&4.58\\
&REKS\_NARM &5.24	&7.13	&9.49	&3.46	&4.06	&4.66\\

\hline 
\end{tabular}\vspace{-2mm}
\end{table}

\textbf{Impact of user information in the KGs.}
In our experiments, Amazon datasets contain user information and item-side information, and thus the constructed KG contains user ID information. To validate the impact of user information, we construct the KGs without user information on Amazon datasets by also deleting the “user purchase product” relationship. Then, we explore the performance of different approaches, while due to space limitation, we only depict the results of NARM and REKS\_NARM.

As shown in Table \ref{tb:userInformation}, without user information, REKS\_NARM still performs better than NARM. That is to say, if the data contains user information, it can be added to the knowledge graph. However, RKES can still work and facilitate the existing non-explainable SR methods without user information in the KGs.


\subsubsection{Sensitivity of Hyper-parameters (RQ3)}
We further study the impact of learning rate $lr$, and $\beta$ in loss function on the model performance. Due to space limitation, we take REKS\_NARM as an example. Figure \ref{fig:hyper} shows the results with the learning rate in the range of $\{0.0001,0.0005,0.001,0.005\}$ and $\beta$ of $\{0.2,0.4,0.6,0.8,1,1.2\}$, respectively. We can see that our framework, although is affected, comparatively insensitive to these hyper-parameters, which helps verify the robustness of our framework.

\subsection{Model Performance on Explanation Task (RQ4)}
In this section, we first introduce the process of generating explanations, and then conduct a survey to evaluate user satisfaction towards the explanations generated by our framework. Finally, we also give three explanation cases generated by REKS\_NARM to intuitively elaborate how REKS interprets the recommendation results. 
\begin{figure}[htbp]
    \centering
    \includegraphics[width=6cm]{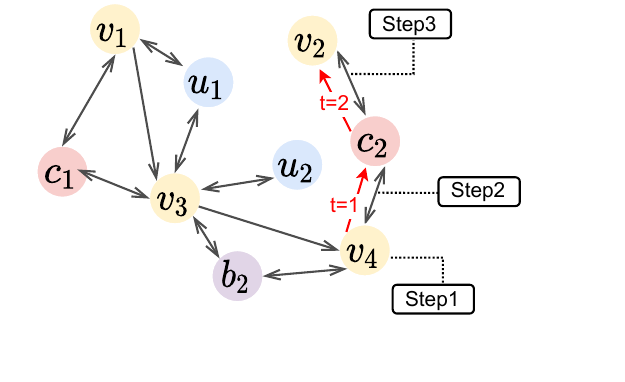}\vspace{-3mm}
    \caption{Generating explanations for session $\{v_1, v_3, v_4\}$.
}\vspace{-3mm}
    \label{fig:explain_model}
\end{figure}

\textbf{Generating explanations.} Specifically, we adopt the probabilistic beam search \cite{xian2019reinforcement} to generate semantic paths on KGs. We denote the predefined sampling sizes at each step as $P_1$ and $P_2$ (path length=$2$). Figure \ref{fig:explain_model} shows the process of generating explanations given session $\{v_1, v_3, v_4\}$, which involves three steps. First, the starting point is determined, where in REKS, the starting point is the last interacted product of the session, i.e., $v_4$. Second, we select the top-$P_1$ first-order neighbors according to the probability $\pi$, where in Figure \ref{fig:explain_model}, one of the first-order neighbors, $c_2$, is selected. Third, 
we determine $P_2$ second-order neighbors with the highest scores (for each first-order neighbor).
Via the three steps, there may exist multiple candidate paths between the starting point ($v_4$) and the predicted entity, such as $v_2$. Then, according to the inner product of the probabilities of steps 2 and 3, we choose the semantic path with the highest overall probability as the one to interpret why we recommend the entity given this session. In our study, $P_1$ and $P_2$ are set to $100$ and $1$, respectively. The semantic path in our scenario is presented in the mode of:
\begin{itemize}
\item $v_4 \xrightarrow{belong\_to}c_2 \xrightarrow{belong\_to} v_2$ 
\end{itemize}

\textbf{User study for verifying the quality of explanations.}
We design a user study\footnote{It is quite challenging to compare our model with other explainable methods in offline settings due to: (1) explainability is difficult to be measured by well-established metrics without ground-truth datasets; (2) to the best of our knowledge, there is no explainable model for SR in the literature; and (3) it is unfair and difficult to compare with other models by an offline user study since the generated recommendations lists may differ a lot.} to validate the suitability and quality of the explanations generated by REKS. In particular, we randomly select $20$ cases (sessions) from the three Amazon datasets (i.e., Baby, Beauty and Cellphones). Each case includes three parts: session information, semantic paths, and the recommended product, where the semantic paths of each case are generated by RKES\_NARM.
Moreover, inspired by \cite{wang2018explainable,park2022reinforcement}, we aim to evaluate explanations from the following six perspectives: \emph{satisfaction}, \emph{effectiveness}, \emph{transparency}, \emph{persuasiveness}, \emph{usability}, and \emph{easy to understand}.
Correspondingly, we design a questionnaire (in a web page) that covers six questions for each case: 
\begin{itemize}
\item \textbf{Satisfaction}: I am satisfied with the explanations.
\item \textbf{Effectiveness}: These explanations help me make more accurate/faster decisions.
\item \textbf{Transparency}: I know why the recommender system recommends this product to me.
\item \textbf{Persuasiveness}: I believe that these explanations can increase the view/purchase rate on the product.
\item \textbf{Unusability}: I am reluctant to use this system.
\item \textbf{Difficult to understand}: I find these explanations difficult
to understand.
\end{itemize}

We recruit $50$ subjects to take participant in the study in an anonymous form, where each subject performs subjective scores on a 1-5 scale (representing strongly disagree, disagree, neutral, agree, and strongly agree, accordingly). Noted that larger values indicate better performance regarding the first four questions, whereas smaller values imply better for the last two. We make a such design (e.g., ask ``unusability'' instead of ``usability'') to control the quality of user study to avoid that some participants of low quality might consistently rate high or low for all questions.

\begin{figure}[t]
    \centering
    \includegraphics[width=7.5cm]{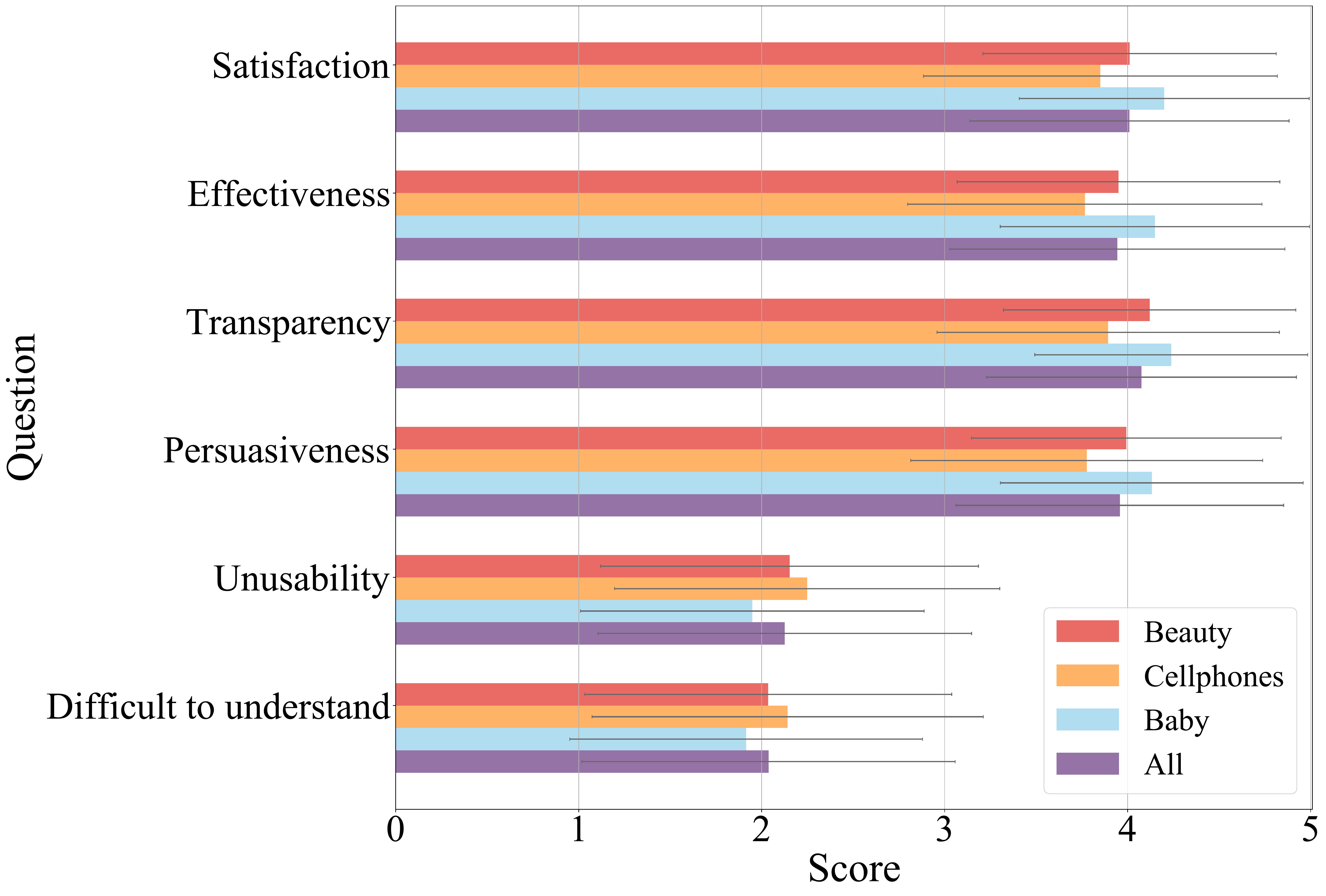}\vspace{-3mm}
    \caption{Questionnaire results of explainability.
}\vspace{-2mm}
    \label{fig:case_score}
\end{figure}

Figure \ref{fig:case_score} displays the mean results from the user study towards the cases from Beauty, Cellphones, Baby, and all the three datasets, respectively, where the horizontal axis means the scores for each perspective, and the black line means the standard deviation. We note that all perspectives are evaluated relatively positively, confirming the effectiveness and rationality of the explanations generated by our REKS.

\begin{figure}[t]
    \centering
    \includegraphics[width=8cm]{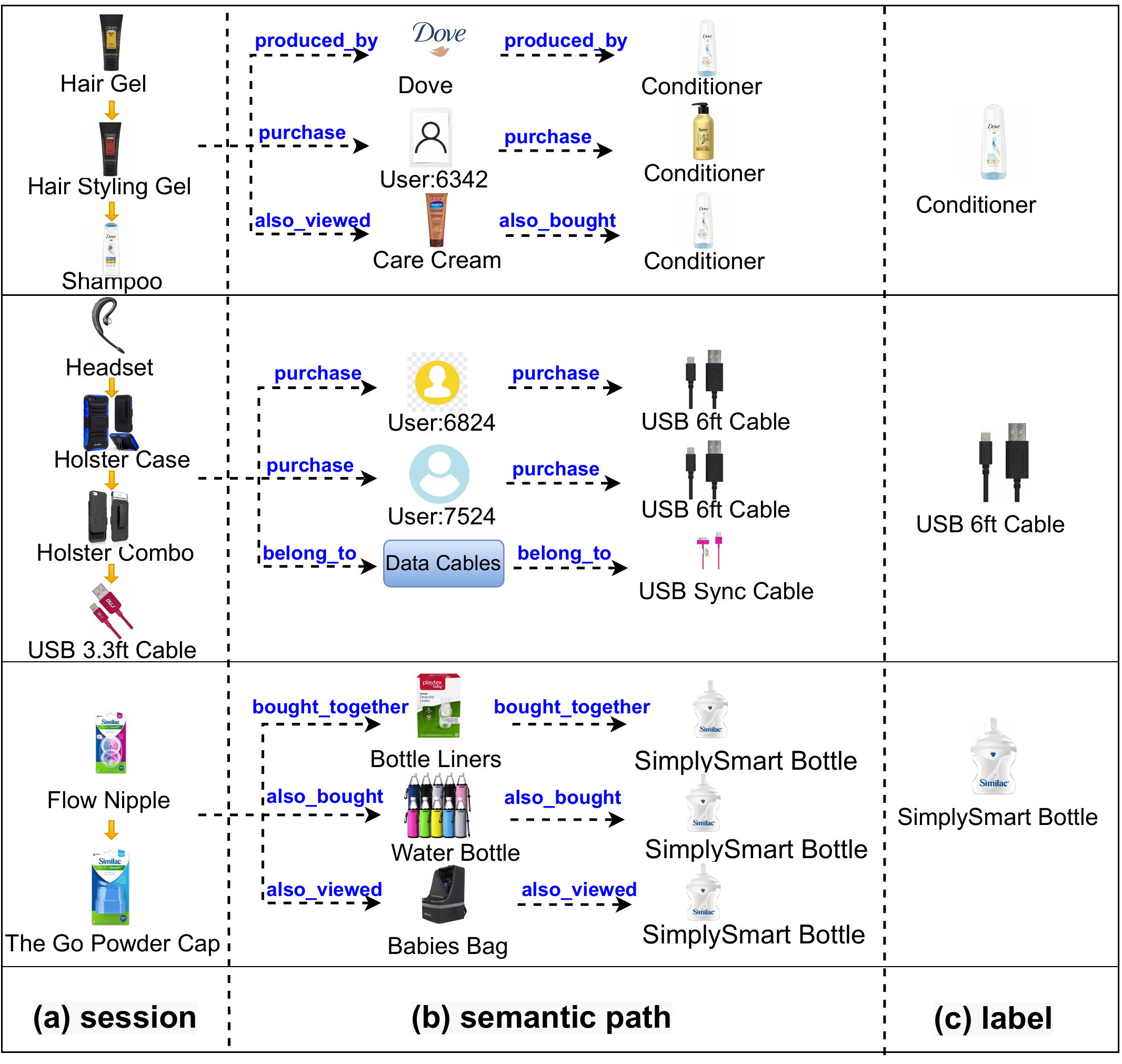}\vspace{-3mm}
    \caption{Semantic paths of three cases.}\vspace{-2mm}
    \label{fig:explainability}
\end{figure}

\textbf{Case study.}
We further showcase three intuitive examples (from three Amazon datasets) to elaborate 
our explanations. Specifically, Figures \ref{fig:explainability} (a) (b), and (c) show session information, three semantic paths of the highest overall probability (as introduced in \textbf{generating explanations} part), and the ground-truth recommended product, respectively.

As shown in the first case on Beauty, given session \{Hair Gel, Hair Styling Gel, Shampoo\}, the recommended product is ``Conditioner''. Three semantic paths are generated, 
where the first and third paths recommend correctly, but the first semantic path is more explanatory, which tells that the last product of the session ``Shampoo'' and the recommended ``Conditioner'' have the same brand ``Dove''. Besides, given the session, we can see that the user's recent interests are related to \emph{hair}, which are consistently implied by all semantic paths.
The second case is from Cellphones dataset, where the session is \{Bluetooth Headset, Holster Case, Holster Combo, USB 3.3ft Cable\}, and the recommended product is ``USB 6ft Cable''. The first two paths provide accurate recommendation, where ``USB 6ft Cable'' is recommended since user 6824/7524 has bought both ``USB 3.3ft Cable'' and ``USB 6ft Cable''. 
Intuitively, there are too many data cables in the same category, merely recommending the products of the same category (as the third semantic path) is less likely to succeed.
For the third case (from Baby), concerning session \{Flow Nipple, The Go Powder Cap\}, ``Simply Smart Bottle'' is recommended since the last product of the session is frequently bought together with ``Bottle Liners'', and ``Bottle Liners''  is frequently bought together with the recommended product. For the session, the user's intent is might about \emph{Baby} and \emph{Bottle}, and ``Bottle Liners'' is the most relevant compared to other entities (e.g., Water Bottle) from the three semantic paths. 

To conclude, via the user study and the three intuitive examples, we can find that REKS can not only generate reasonable explanations, but also can better capture real intentions implied by every session. Besides, our design of using the last product of a session as the starting point of semantic paths is also verified to be effective.

\section{Conclusions}
\label{sec:concl}
Existing representative session-based recommendation, especially deep learning-based, often acts as a ``black box'' and cannot provide explanations, leading to lowered user satisfaction and acceptance. In this paper, we design a generic reinforced explainable framework with knowledge graph (denoted as REKS) to simultaneously address recommendation and explanation tasks for session-based recommendation. REKS can be easily and flexibly applied into non-explainable SR methods to facilitate the explanation task by generating path-based explanations from knowledge graph. In particular, REKS constructs a knowledge graph with session information and treats SR models as part of the policy network of Markov decision process with particularly designed state vector, reward strategy and loss function. By instantiating the REKS in five representative SR models (i.e., GRU4REC, NARM, SR-GNN, GCSAN, and BERT4REC), experimental results on
four datasets demonstrate the effectiveness of our framework on both recommendation and explanation tasks, and the rationality of our model designs.
For future study, we strive to design alternative ways, e.g., A/B test and online experiments, to validate the superiority of our path-based explainable framework over other explainable approaches.

\section*{Acknowledgment}
This work was supported in part by the National Natural Science Foundation of China (Grant No. 72192832), the Shanghai Natural Science Foundation of China (Grant No. 21ZR1421900), and Ant Group. It was also supported by A*Star
Center for Frontier Artificial Intelligence Research and in
part by the Data Science and Artificial Intelligence Research Centre, School of Computer Science and Engineering at
the Nanyang Technological University (NTU), Singapore.


\newpage

\bibliographystyle{IEEEtran}
\bibliography{REKS}

\begin{thebibliography}{10}
\providecommand{\url}[1]{#1}
\csname url@samestyle\endcsname
\providecommand{\newblock}{\relax}
\providecommand{\bibinfo}[2]{#2}
\providecommand{\BIBentrySTDinterwordspacing}{\spaceskip=0pt\relax}
\providecommand{\BIBentryALTinterwordstretchfactor}{4}
\providecommand{\BIBentryALTinterwordspacing}{\spaceskip=\fontdimen2\font plus
\BIBentryALTinterwordstretchfactor\fontdimen3\font minus
  \fontdimen4\font\relax}
\providecommand{\BIBforeignlanguage}[2]{{%
\expandafter\ifx\csname l@#1\endcsname\relax
\typeout{** WARNING: IEEEtran.bst: No hyphenation pattern has been}%
\typeout{** loaded for the language `#1'. Using the pattern for}%
\typeout{** the default language instead.}%
\else
\language=\csname l@#1\endcsname
\fi
#2}}
\providecommand{\BIBdecl}{\relax}
\BIBdecl

\bibitem{wang2021survey}
S.~Wang, L.~Cao, Y.~Wang, Q.~Z. Sheng, M.~A. Orgun, and D.~Lian, ``A survey on
  session-based recommender systems,'' \emph{ACM Computing Surveys (CSUR)},
  vol.~54, no.~7, pp. 1--38, 2021.

\bibitem{le2016modeling}
D.-T. Le, Y.~Fang, and H.~W. Lauw, ``Modeling sequential preferences with
  dynamic user and context factors,'' in \emph{Proceedings of the Joint
  European Conference on Machine Learning and Knowledge Discovery in
  Databases}.\hskip 1em plus 0.5em minus 0.4em\relax Springer, 2016, pp.
  145--161.

\bibitem{li2017neural}
J.~Li, P.~Ren, Z.~Chen, Z.~Ren, T.~Lian, and J.~Ma, ``Neural attentive
  session-based recommendation,'' in \emph{Proceedings of the 26th ACM
  International Conference on Conference on Information and Knowledge
  Management}, 2017, pp. 1419--1428.

\bibitem{sun2019bert4rec}
F.~Sun, J.~Liu, J.~Wu, C.~Pei, X.~Lin, W.~Ou, and P.~Jiang, ``Bert4rec:
  Sequential recommendation with bidirectional encoder representations from
  transformer,'' in \emph{Proceedings of the 28th ACM International Conference
  on Information and Knowledge Management}, 2019, pp. 1441--1450.

\bibitem{wang2020global}
Z.~Wang, W.~Wei, G.~Cong, X.-L. Li, X.-L. Mao, and M.~Qiu, ``Global context
  enhanced graph neural networks for session-based recommendation,'' in
  \emph{Proceedings of the 43rd International ACM SIGIR Conference on Research
  and Development in Information Retrieval}, 2020, pp. 169--178.

\bibitem{pang2022heterogeneous}
Y.~Pang, L.~Wu, Q.~Shen, Y.~Zhang, Z.~Wei, F.~Xu, E.~Chang, B.~Long, and
  J.~Pei, ``Heterogeneous global graph neural networks for personalized
  session-based recommendation,'' in \emph{Proceedings of the 15th ACM
  International Conference on Web Search and Data Mining}, 2022, pp. 775--783.

\bibitem{wang2018ripplenet}
H.~Wang, F.~Zhang, J.~Wang, M.~Zhao, W.~Li, X.~Xie, and M.~Guo, ``Ripplenet:
  Propagating user preferences on the knowledge graph for recommender
  systems,'' in \emph{Proceedings of the 27th ACM International Conference on
  Information and Knowledge Management}, 2018, pp. 417--426.

\bibitem{ai2018learning}
Q.~Ai, V.~Azizi, X.~Chen, and Y.~Zhang, ``Learning heterogeneous knowledge base
  embeddings for explainable recommendation,'' \emph{Algorithms}, vol.~11,
  no.~9, p. 137, 2018.

\bibitem{wang2019explainable}
X.~Wang, D.~Wang, C.~Xu, X.~He, Y.~Cao, and T.-S. Chua, ``Explainable reasoning
  over knowledge graphs for recommendation,'' in \emph{Proceedings of the 33rd
  AAAI Conference on Artificial Intelligence}, vol.~33, no.~01, 2019, pp.
  5329--5336.

\bibitem{xian2020cafe}
Y.~Xian, Z.~Fu, H.~Zhao, Y.~Ge, X.~Chen, Q.~Huang, S.~Geng, Z.~Qin, G.~De~Melo,
  S.~Muthukrishnan \emph{et~al.}, ``Cafe: Coarse-to-fine neural symbolic
  reasoning for explainable recommendation,'' in \emph{Proceedings of the 29th
  ACM International Conference on Information and Knowledge Management}, 2020,
  pp. 1645--1654.

\bibitem{zhao2020leveraging}
K.~Zhao, X.~Wang, Y.~Zhang, L.~Zhao, Z.~Liu, C.~Xing, and X.~Xie, ``Leveraging
  demonstrations for reinforcement recommendation reasoning over knowledge
  graphs,'' in \emph{Proceedings of the 43rd International ACM SIGIR Conference
  on Research and Development in Information Retrieval}, 2020, pp. 239--248.

\bibitem{huang2019explainable}
X.~Huang, Q.~Fang, S.~Qian, J.~Sang, Y.~Li, and C.~Xu, ``Explainable
  interaction-driven user modeling over knowledge graph for sequential
  recommendation,'' in \emph{Proceedings of the 27th ACM International
  Conference on Multimedia}, 2019, pp. 548--556.

\bibitem{lin2018multi}
X.~V. Lin, C.~Xiong, and R.~Socher, ``Multi-hop knowledge graph reasoning with
  reward shaping,'' in \emph{Proceedings of the 2018 Conference on Empirical
  Methods in Natural Language Processing}, 2018.

\bibitem{geng2022path}
S.~Geng, Z.~Fu, J.~Tan, Y.~Ge, G.~De~Melo, and Y.~Zhang, ``Path language
  modeling over knowledge graphs for explainable recommendation,'' in
  \emph{Proceedings of the ACM Web Conference 2022}, 2022, pp. 946--955.

\bibitem{kamehkhosh2017comparison}
I.~Kamehkhosh, D.~Jannach, and M.~Ludewig, ``A comparison of frequent pattern
  techniques and a deep learning method for session-based recommendation.'' in
  \emph{Proceedings of the 11th ACM Conference on Recommender Systems}, 2017,
  pp. 50--56.

\bibitem{shani2005mdp}
G.~Shani, D.~Heckerman, R.~I. Brafman, and C.~Boutilier, ``An mdp-based
  recommender system.'' \emph{Journal of Machine Learning Research}, vol.~6,
  no.~9, 2005.

\bibitem{rendle2010factorizing}
S.~Rendle, C.~Freudenthaler, and L.~Schmidt-Thieme, ``Factorizing personalized
  markov chains for next-basket recommendation,'' in \emph{Proceedings of the
  19th International Conference on World Wide Web}, 2010, pp. 811--820.

\bibitem{he2016fusing}
R.~He and J.~McAuley, ``Fusing similarity models with markov chains for sparse
  sequential recommendation,'' in \emph{IEEE International Conference on Data
  Mining}.\hskip 1em plus 0.5em minus 0.4em\relax IEEE, 2016, pp. 191--200.

\bibitem{he2017translation}
R.~He, W.-C. Kang, and J.~McAuley, ``Translation-based recommendation,'' in
  \emph{Proceedings of the 11th ACM Conference on Recommender Systems}, 2017,
  pp. 161--169.

\bibitem{hidasi2015session}
B.~Hidasi, A.~Karatzoglou, L.~Baltrunas, and D.~Tikk, ``Session-based
  recommendations with recurrent neural networks,'' in \emph{Proceedings of the
  4th International Conference on Learning Representations}, 2016.

\bibitem{chen2020handling}
T.~Chen and R.~C.-W. Wong, ``Handling information loss of graph neural networks
  for session-based recommendation,'' in \emph{Proceedings of the 26th ACM
  SIGKDD International Conference on Knowledge Discovery and Data Mining},
  2020, pp. 1172--1180.

\bibitem{wu2019session}
S.~Wu, Y.~Tang, Y.~Zhu, L.~Wang, X.~Xie, and T.~Tan, ``Session-based
  recommendation with graph neural networks,'' in \emph{Proceedings of the 33rd
  AAAI Conference on Artificial Intelligence}, vol.~33, no.~01, 2019, pp.
  346--353.

\bibitem{li2015gated}
Y.~Li, D.~Tarlow, M.~Brockschmidt, and R.~Zemel, ``Gated graph sequence neural
  networks,'' in \emph{Proceedings of the 3rd International Conference on
  Learning Representations}, 2015.

\bibitem{qiu2019rethinking}
R.~Qiu, J.~Li, Z.~Huang, and H.~Yin, ``Rethinking the item order in
  session-based recommendation with graph neural networks,'' in
  \emph{Proceedings of the 28th ACM International Conference on Information and
  Knowledge Management}, 2019, pp. 579--588.

\bibitem{xu2019graph}
C.~Xu, P.~Zhao, Y.~Liu, V.~S. Sheng, J.~Xu, F.~Zhuang, J.~Fang, and X.~Zhou,
  ``Graph contextualized self-attention network for session-based
  recommendation,'' in \emph{Proceedings of the 28th International Joint
  Conference on Artificial Intelligence}, vol.~19, 2019, pp. 3940--3946.

\bibitem{zhang2020explainable}
Y.~Zhang, X.~Chen \emph{et~al.}, ``Explainable recommendation: A survey and new
  perspectives,'' \emph{Foundations and Trends{\textregistered} in Information
  Retrieval}, vol.~14, no.~1, pp. 1--101, 2020.

\bibitem{zhang2014explicit}
Y.~Zhang, G.~Lai, M.~Zhang, Y.~Zhang, Y.~Liu, and S.~Ma, ``Explicit factor
  models for explainable recommendation based on phrase-level sentiment
  analysis,'' in \emph{Proceedings of the 37th International ACM SIGIR
  conference on Research and Development in Information Retrieval}, 2014, pp.
  83--92.

\bibitem{hou2019explainable}
H.~Hou and C.~Shi, ``Explainable sequential recommendation using knowledge
  graphs,'' in \emph{Proceedings of the 5th International Conference on
  Frontiers of Educational Technologies}, 2019, pp. 53--57.

\bibitem{cui2021reinforced}
Z.~Cui, H.~Chen, L.~Cui, S.~Liu, X.~Liu, G.~Xu, and H.~Yin, ``Reinforced kgs
  reasoning for explainable sequential recommendation,'' \emph{World Wide Web},
  pp. 1--24, 2021.

\bibitem{chen2021ssr}
J.~Chen, W.~Wu, W.~Hu, W.~Zheng, and L.~He, ``Ssr: Explainable session-based
  recommendation,'' in \emph{Proceedings of the 2021 International Joint
  Conference on Neural Networks}.\hskip 1em plus 0.5em minus 0.4em\relax IEEE,
  2021, pp. 1--8.

\bibitem{geng2022causality}
C.~Geng, H.~Wu, and H.~Fang, ``Causality and correlation graph modeling for
  effective and explainable session-based recommendation,'' \emph{arXiv
  preprint arXiv:2201.10782}, 2022.

\bibitem{bordes2013translating}
A.~Bordes, N.~Usunier, A.~Garcia-Duran, J.~Weston, and O.~Yakhnenko,
  ``Translating embeddings for modeling multi-relational data,'' \emph{Advances
  in Neural Information Processing Systems}, vol.~26, 2013.

\bibitem{xian2019reinforcement}
Y.~Xian, Z.~Fu, S.~Muthukrishnan, G.~De~Melo, and Y.~Zhang, ``Reinforcement
  knowledge graph reasoning for explainable recommendation,'' in
  \emph{Proceedings of the 42nd International ACM SIGIR Conference on Research
  and Development in Information Retrieval}, 2019, pp. 285--294.

\bibitem{balloccu2022post}
G.~Balloccu, L.~Boratto, G.~Fenu, and M.~Marras, ``Post processing recommender
  systems with knowledge graphs for recency, popularity, and diversity of
  explanations,'' in \emph{Proceedings of the 45th International ACM SIGIR
  Conference on Research and Development in Information Retrieval}, 2022.

\bibitem{sutton2018reinforcement}
R.~S. Sutton and A.~G. Barto, \emph{Reinforcement learning: An
  introduction}.\hskip 1em plus 0.5em minus 0.4em\relax MIT Press, 2018.

\bibitem{wang2019multi}
H.~Wang, F.~Zhang, M.~Zhao, W.~Li, X.~Xie, and M.~Guo, ``Multi-task feature
  learning for knowledge graph enhanced recommendation,'' in \emph{Proceedings
  of the International Conference on World Wide Web 2019}, 2019, pp.
  2000--2010.

\bibitem{wang2018explainable}
N.~Wang, H.~Wang, Y.~Jia, and Y.~Yin, ``Explainable recommendation via
  multi-task learning in opinionated text data,'' in \emph{Proceedings of the
  41st International ACM SIGIR Conference on Research and Development in
  Information Retrieval}, 2018, pp. 165--174.

\bibitem{park2022reinforcement}
S.-J. Park, D.-K. Chae, H.-K. Bae, S.~Park, and S.-W. Kim, ``Reinforcement
  learning over sentiment-augmented knowledge graphs towards accurate and
  explainable recommendation,'' in \emph{Proceedings of the 15th ACM
  International Conference on Web Search and Data Mining}, 2022, pp. 784--793.

\end{thebibliography}

\end{document}